\documentclass[12pt,a4paper]{article}
\usepackage{amssymb,amsfonts}
\usepackage[centertags]{amsmath}
\usepackage{graphicx,psfrag}
\usepackage{color}
\newcommand{\be}{\begin{equation}}
\newcommand{\ee}{\end{equation}}
\newcommand{\bea}{\begin{eqnarray}}
\newcommand{\eea}{\end{eqnarray}}
\newcommand{\bdm}{\begin{displaymath}}
\newcommand{\edm}{\end{displaymath}}
\newcommand{\lb}{\label}

\begin{document}
\begin{titlepage}
\begin{flushright}
RESCEU-23/10
\end{flushright}
\vskip 0.7cm
\begin{center}
{\large\bf  COSMOLOGICAL CONSTANT FROM DECOHERENCE}
\vskip 1cm
{\bf Claus Kiefer and Friedemann Queisser}
\vskip 0.4cm
 Institut f\"ur Theoretische Physik, Universit\"at zu K\"oln,\\
 Z\"ulpicher Strasse~77, 50937 K\"oln, Germany.
\vskip 0.7cm
{\bf Alexei A. Starobinsky}
\vskip 0.4cm
 Landau Institute for Theoretical Physics,
 Moscow 119334, Russia and \\
 RESCEU, Graduate School of Science, The University of Tokyo,
 Tokyo 113-0033, Japan
\end{center}
\date{\today}
\vskip 2cm
\begin{center}
{\bf Abstract}
\end{center}
\small
\begin{quote}
We address the issue why a cosmological constant (dark energy)
possesses a small positive value instead of being zero. 
Motivated by the cosmic landscape picture, we mimic the dark energy by
a scalar field with potential wells and show that other degrees
of freedom interacting with it can localize this field by decoherence
in one of the wells. Dark energy
can then acquire a small positive value. We also show that the additional
degrees of freedom enhance the tunneling rate between the wells. The
consideration is performed in detail for the case of two wells and then
extended to a large number of wells.  
\end{quote}
\normalsize
\end{titlepage}
\section{Introduction}
Observations indicate that our Universe is currently accelerating
\cite{WMAP7}. The simplest explanation of this fact in terms of the 
Einstein equations of gravity is the existence of a non-zero Einstein 
cosmological constant $\Lambda$. However, a more general dynamical entity 
dubbed ``dark energy'' (DE), whose effective energy--momentum tensor is 
close to that of $\Lambda$, is also a serious possibility 
\cite{sahnistaro,pad,peebles,CST06,SS06,FTH}. Present measurements of the
effective DE equation of state $w_{\rm DE}:= p_{\rm DE}/\rho_{\rm DE}$ are 
inconclusive about its nature and are compatible with DE being 
given exactly by $\Lambda$, that is, with $w_{\rm DE}\equiv -1$
\cite{WMAP7}. 
 
The microscopic origin of DE is presently unknown and can perhaps
only be explained after a full quantum theory of gravity is available
\cite{OUP}. There are actually two aspects connected with DE.
First, it is unclear why the effective energy density of DE (or $\Lambda$
itself) is much smaller than one would expect from elementary 
particle physics on dimensional grounds. Second, it is unclear why it is
not exactly zero but has a small positive value. The first question
might only be answered on the basis of a full quantum theory of
all quantum fields including gravity. But it is possible that the second 
question can be addressed on the basis of known physics.

A first step in the understanding of the second question
 was made by Yokoyama \cite{yoko}.
He modelled DE by a scalar field $\phi$ with a potential
$V(\phi)$ that is characterized by a double well. This choice can be
motivated by recent ideas in string theory, where a
``landscape'' of many (perhaps as many as $10^{500}$ or more) local minima of a
complicated potential is discussed, see, for example, \cite{carr,Douglas} and
the references therein.  
Yokoyama assumed that, perhaps due to some unknown symmetry, the exact
ground state of the Universe is characterized by a vanishing vacuum
energy, that is, a vanishing cosmological constant and that
the observed small deviation from zero could arise from the 
fact that the Universe is not in its ground state.   

But how can this happen?
The ground state for a double-well potential is extended (delocalized)
over both minima. In contrast to this, a state localized in one of the
minima is a {\em superposition} of the eigenstates; in the simplest case, it
is a superposition of the ground and the first excited state. 
The effective energy of such localized states is greater than the
ground-state energy and would therefore be positive if the
ground-state energy were zero. If the wall between the 
wells is not too small, the values for this positive-energy states are tiny
because they differ from the ground-state energy only by a small
tunneling factor proportional to $\exp(-S_0)$, where $S_0$ is the instanton
action. The reason for the observed small positive cosmological
constant could thus lie in the fact that the dark energy in the Universe is in a
localized state that is concentrated near one of the minima of the
potential. 

As long as the universe stays in such a localized state, the effective DE
equation of state will be $w_{\rm DE}\approx -1$. There exists,
however, a certain 
probability that the Universe can evolve into the ground state, which
is a superposition of the localized states, or tunnel into another
localized state.
The question then arises how large the time scale and
the tunneling rate would be. Obviously, 
these scales should obey all known observational constraints
presented in \cite{WMAP7}. 

In our paper we shall elaborate on this idea in two respects. 
First, 
it has to be justified why a DE field is not found in its
ground state in the first place, but in 
a localized state.  This is reminiscent of an analogous situation in
quantum mechanics where chiral molecules are typically found in
left-handed or right-handed isomeric forms and not in a superposition
of the two (as e.g. the ground state would be).  Historically, this
was called ``Hund's paradox'' \cite{deco}. Its solution is based on
the central concept of decoherence \cite{deco,schlosshauer} and was
recently presented for a specific case in quantitative detail in
\cite{TK09}.  Decoherence is 
the unavoidable and irreversible emergence of classical properties
with its environment; ``environment'' stands here for any irrelevant degrees
of freedom that interact with the quantum system and thereby become entangled
with it. In the case of the chiral molecules these can be photons or
air molecules scattering off these molecules.  
Decoherence will also play a central role in our analysis, 
in order to explain why the
DE field is not in its delocalized ground state, but in a
localized state with higher energy.
 
Our second elaboration is a direct consequence of the first one. If
additional ``decohering'' degrees of freedom are present, they will have
an effect on the tunneling rate from one well to the other.  
We shall thus discuss both the pure tunneling rate
of the isolated system as well as its modification by the
environment. In view
of the experience from quantum mechanical models \cite{caldlegg}, one
would expect that these degrees of freedom 
will in general {\em reduce} 
this rate, so that tunneling will become less likely.
In contrast to this expectation, however,
we shall find that this tunneling rate is here
increased rather than decreased, by a field-theoretic mechanism similar
to the Casimir effect. 
The localization by the environment together with the magnitude of the
tunneling rate could then provide the explanation of why we observe a small
positive value for dark energy.  

We note that our approach is completely different
from the one used in the paper \cite{CL10} (which appeared
when our paper was prepared for publication), where entanglement
between different cosmological epochs, not different vacua,
is proposed as a source of $\Lambda$.

Our considerations should also be relevant for the
inflationary stage in the early Universe, which was dominated by
a primordial DE whose properties were close to an effective 
(though metastable) cosmological constant, too. However, we do not 
attempt here an application in this direction. 

Our paper is organized as follows. In Sec.~2.1 we present our model:
DE is modelled by a scalar field (quintessence), and it interacts with
an environmental massless scalar field that may arise from metric
perturbations. In Sec.~2.2 the process of decoherence is discussed in
detail for both sub- and super-Hubble modes of the environment. It is
found that these modes localize the DE field in one well, the
super-Hubble modes even more efficiently than the sub-Hubble modes. 
In Sec.~3 we calculate the tunneling rate for a Minkowski background
as well as an expanding Universe. We find that the presence of an
environmental field {\em enhances} the original tunneling rate. 
In order to approach more closely the ideas of a
cosmic landscape, we extend our discussion in Sec.~4 to the presence
of many vacua. A brief Appendix summarizes some material about the
$\zeta$-function renormalization needed in Sec.~3.

Our main results have already been briefly announced in \cite{MG12}.
\section{Small positive dark energy from decoherence}
\subsection{The model}
We present here our model in which a scalar field $\phi$ mimicking DE
(called the ``system'' in the following) is coupled  to other degrees
of freedom called ``environment''. As in the Yokoyama paper \cite{yoko},
$\phi$ is assumed to possess a potential with two quasi-localized
minima.  As for the environment, we choose another  scalar field
 called $\sigma$ which
couples to $\phi$.  The details of the interaction are of secondary
importance; it only has to be able to generate a sufficient
entanglement between system and environment \cite{deco}. 

Both system and environment are supposed to evolve in a flat 
Friedmann-Robertson-Walker (FRW) background
with the line element 
\begin{eqnarray}
\lb{FRW}
ds^2=g^{\rm B}_{\mu\nu}dx^\mu dx^\nu= d t^2 - a^2(t)(dx^2+dy^2+dz^2)
\end{eqnarray}
($\hbar=c=1$ is set throughout the paper).
We assume here for later convenience that the
scale factor, $a$, has the dimension of a length, while $x$, $y$, and
$z$ are dimensionless.  
The total action of system and environment then reads
\begin{eqnarray}\label{22}
S=\int d^4x
\sqrt{-g^\mathrm{B}}\left(\mathcal{L}_{\rm sys}+\mathcal{L}_{\rm
    env}+\mathcal{L}_{\rm int}\right)\,, 
\end{eqnarray}
where $g^\mathrm{B}$ is the determinant of the metric $g^{\rm B}_{\mu\nu}$,
and $\mathcal{L}_{\rm sys}$, $\mathcal{L}_{\rm env}$, and
$\mathcal{L}_{\rm int}$ denote 
the Lagrangian of the system, environment, and interaction, respectively.
The spatially homogeneous scalar field describing the vacuum energy reads
\begin{eqnarray}
\label{0}
\mathcal{L}_{\rm sys}:=\frac{1}{2} \dot{\phi}^2 -V(\phi)\, ,
\end{eqnarray}
while
the Lagrangian of the environment, $\mathcal{L}_{\rm env}$,
describes a massless scalar field $\sigma$,
\begin{eqnarray}
\lb{envsigma}
\mathcal{L}_{{\rm env},\sigma}:=
\frac{1}{2}g^{\rm B\mu\nu}\partial_\mu \sigma\partial_\nu \sigma\,.
\end{eqnarray}

The interaction between system and environment
must be able to discriminate between different values of the scalar
field $\phi$. 
In quantum mechanical applications one often chooses a \textit{bilinear}
interaction, 
see for instance the Caldeira--Leggett model \cite{caldlegg} or the
spin--boson model described, for example,
in Section~5.3 of \cite{schlosshauer}. Since, however, our
field $\phi$ is spatially \textit{homogeneous},
a bilinear term  of the form $\int d^3x\ \sigma(\mathbf{x},t)\phi(t)$
would give no interesting dynamics, since only a single Fourier component of the
scalar field $\sigma$, the one with vanishing momenta, would interact with
the system field. 
We therefore introduce the \textit{tri-linear} interaction
\begin{eqnarray}
\label{trilinear}
\mathcal{L}_{{\rm int},\sigma}:=-g_{\rm s} \sigma^2\phi\, 
\end{eqnarray}
for the coupling with the scalar-field environment.
Here, the coupling constant $g_{\rm s}$ has to be chosen
such that the product $g_{\rm s}\phi$ is positive.
This guarantees that the corresponding Hamiltonian is
bounded from below and that the dynamics is thus stable.
Since both $\sigma^2$ and $\phi^2$ have physical dimension mass ($M$) over
length ($L$), the dimension of $g_{\rm s}$ is $(ML^3)^{-1/2}$, which
in the natural units used here is equal to $M$.

What could be the origin of such a coupling?
It can arise, for example, from the expansion of the
metric determinant $\sqrt{-g}$ into the scalar and tensor modes.
Let us consider therefore the FRW line element with scalar
and tensor perturbations (see e.g. \cite{Mukhanov}),
\begin{eqnarray}
ds^2=(1+2\psi_1)dt^2-a^2\left[(1-2\psi_2)\delta_{ij}+h_{ij}\right]dx^idx^j\,,
\end{eqnarray}
where $\psi_{1/2}$ are scalar perturbations and $h_{ij}$ are 
tensor modes.
In the transverse and traceless gauge, there are only two independent
tensor modes, $h_{23}=h_{32}$ and $h_{22}=-h_{33}$.
If there is a linear term in the potential of the system field $\phi$,
we get from the expansion of the determinant a term of the form
\begin{eqnarray}
& & \int d^3 x\sqrt{-g}\phi\approx\\ & &  \int d^3
x\sqrt{-g^B}\left(1+\psi_1-3\psi_2-2\psi_1^2+
\frac{3}{2}(\psi_1-\psi_2)^2-\frac{1}{2}h_{22}^2-\frac{1}{2}h_{23}^2\right)\phi\,.
\nonumber
\end{eqnarray}
Discarding the terms linear in $\psi_1$ and $\psi_2$ for the reason
mentioned above, we are left with tri-linear interactions of the
form (\ref{trilinear}); the role of our field $\sigma$ could thus be
played by $\psi_1$, $\psi_2$, $h_{22}$, or $h_{23}$. Such interactions
can thus arise both from the 
scalar and the tensor perturbations of the metric itself, although it
is conceivable that $\sigma$ is any other field that occurs in a
fundamental theory. 

We shall now address the Hamiltonian that will be used in the quantum theory.
It can be derived from (\ref{22})--(\ref{trilinear}) and in the momentum 
representation reads
\begin{eqnarray}\label{20}
H_{\phi,\sigma}&=&\int dp^3\Bigg(\frac{1}{2
  a^3}\tilde{\Pi}(\mathbf{p})\tilde{\Pi}(-\mathbf{p})+\frac{a}{2}p^2\tilde{\sigma}(\mathbf{p})
\tilde{\sigma}(-\mathbf{p})\\   
& &+g_{\rm s} a^3\phi\tilde{\sigma}(\mathbf{p})\tilde{\sigma}(-\mathbf{p})
\Bigg)+H_\phi\nonumber\, ,
\end{eqnarray}
where 
\begin{eqnarray}
\sigma(\mathbf{x})&=&\frac{1}{(2\pi)^{3/2}}\int d^3p\
\tilde{\sigma}_{ab}(\mathbf{p})e^{i\mathbf{x}\mathbf{p}}\,,\\
\Pi(\mathbf{x})&=&\frac{1}{(2\pi)^{3/2}}\int d^3p\
\tilde{\Pi}(\mathbf{p})e^{i\mathbf{x}\mathbf{p}}\, ,
\end{eqnarray}
and $H_\phi$ denotes the pure $\phi$-Hamiltonian (see \eqref{Hphi} below). 
Note that $\mathbf{p}$ is dimensionless because $\mathbf{x}$ is
dimensionless. 

In the following, we shall simplify the part of the
Hamiltonian describing the system.
We shall assume that the dynamics of the system is dominated by the
two lowest energy eigenvalues of the double-well potential, that is,
we assume that
their difference is much smaller than the energy gaps within a single
well.
It is then possible to reduce the system to an {\em effective two-state system},
\begin{eqnarray}
\label{Hphi}
H_\phi=\begin{pmatrix} E_+ & \Delta \\ \Delta & E_-  \end{pmatrix}\, ,
\end{eqnarray}
where $E_+$ and $E_-$ are the energy levels of the localized minima, and
$\Delta$ is the tunneling matrix element. 

The reduction of the system to an effective two-state system leads to the
interaction
\begin{eqnarray}\label{intsigma}
H_{{\rm int},\sigma}=g_{\rm s} a^3(t) \begin{pmatrix} \phi_+  & 0 \\ 0 & \phi_-
\end{pmatrix}\int d^3p\ \sigma(\mathbf{p})\sigma(-\mathbf{p})\,,
\end{eqnarray}
where the environment can only discriminate between the two different minima
of the potential, $\phi_+$ and $\phi_-$.

All together, our model resembles a spin--boson model
\cite{schlosshauer}, although the 
coupling in the standard situation is taken to be linear in the
environmental fields. It is well known that situations with a
double-well potential can often be described by an effective two-state
system \cite{deco,schlosshauer}. In this context the Hamiltonian
\eqref{Hphi} would be written in the form
\bdm
H_\phi=\sigma_x\Delta+\frac12(\delta
E)\sigma_z+\frac12(E_++E_-){\mathbb I}\ ,
\edm
where $\sigma_x$ and $\sigma_z$ are Pauli matrices, and $\delta
E:=E_+-E_-$. 
Spin--boson models describe the
interaction of a  
central system with its environment in the case when the system is
effectively acting as a two-level system.

\subsection{The reduced density matrix}

With this simplification at hand, it is possible to calculate the reduced
density matrix of the two-state system. 
We assume that the initial state is a product of a system and an
environmental state,
$|\Psi\rangle_\mathrm{sys}\otimes|\Psi\rangle_\mathrm{env}$.
The time evolution will then generate an entanglement between them. 
This evolution is governed by the functional Schr\"odinger equation
\begin{equation}
i|\dot\Psi\rangle=H_\mathrm{\phi,\sigma}|\Psi\rangle\ .
\end{equation}
In this section, we shall neglect tunneling, that is, we set
 $\Delta=0$ in \eqref{Hphi}; this enables us 
to solve the Schr\"odinger equation exactly.
We assume that the state of system and environment is of Gaussian
form (see e.g. \cite{kuchar,QED,Jackiw}) and make the ansatz
\begin{eqnarray}
|\Psi\rangle=\begin{pmatrix}\alpha N_+(t)\exp\left(-\frac{1}{2}\int d^3p \
    \sigma(\mathbf{p})\Omega_+(\phi,p,t)\sigma(-\mathbf{p})-i E_+ t\right)\\ 
{} \\ \beta N_-(t)\exp\left(-\frac{1}{2}\int d^3p \
  \sigma(\mathbf{p})\Omega_-(\phi,p,t)\sigma(-\mathbf{p})-i E_-
  t\right)\end{pmatrix}\, ,
\end{eqnarray}
where $\Omega_{+/-}(p,t)$ and $N_{+/-}(t)$ are time-dependent
functions to be determined from the 
Schr\"odinger equation, and $\alpha$ and $\beta$ are constants with
$|\alpha|^2+|\beta|^2=1$. 
With the above ansatz one obtains the
following Riccati-type equations, cf. \cite{QED}, 
\begin{eqnarray}\label{scalequ}
-i\dot{\Omega}_{+/-}(p,t)=-\frac{(\Omega_{+/-}(p,t))^2}{a^3}+p^2
a+g_{\rm s}a^3\phi_{+/-} 
\end{eqnarray}
and
\begin{eqnarray}
\frac{1}{2
  a^3}\mathrm{Tr}\,\Omega_{+/-}(p,t)=i\frac{\dot{N}_{+/-}}{N_{+/-}}\,
. 
\end{eqnarray}
Eq. \eqref{scalequ} can be transformed by the ansatz
\begin{eqnarray}
\Omega_{+/-}(p,t)=:-i a^3\frac{\dot{u}_{+/-}(p)}{u_{+/-}(p)}
\end{eqnarray}
into a linear equation for ${u}_{+/-}$.
Switching to conformal time, which is defined by $a\,d\eta =dt$,
and denoting derivatives with respect to $\eta$ by a prime,  we obtain
\begin{eqnarray}\label{scaleequ2}
(p^2+\gamma
a^2)u_{+/-}(p,\eta)+\frac{2a'}{a}{u'}_{+/-}(p,\eta)+{u''}_{+/-}(p,\eta)=0\,,  
\end{eqnarray}
where we have introduced the quantity
\be
\lb{gamma}
\gamma:=2g_{\rm s}\phi_{+/-}\ ,
\ee
which has dimension $L^{-2}\equiv M^2$ and obeys $\gamma>0$.
(We have skipped the indices $+/-$ in $\gamma$ for simplicity.) We
assume that the coupling between system and environment is small, that
is, $\gamma < H^2$.
The total density matrix for system and environment is given by the pure state
\begin{eqnarray}
\rho(t)=|\Psi\rangle\langle \Psi|\ ,
\end{eqnarray}
from which the reduced density matrix is obtained by integrating out
the environmental scalar field,
\begin{eqnarray}
\rho(t)_\mathrm{sys}:=\mathrm{Tr}_\mathrm{env}|\Psi\rangle\langle \Psi|\,.
\end{eqnarray}
In position representation, this reads
\be
\rho_{ij}=\int {\mathcal D}\sigma\ \Psi_i^*[\sigma]\Psi_j[\sigma]\ ,
\ee
where $i,j$ run over the values $+$ and $-$, and $\Psi\equiv (\psi_+\
\psi_-)^{\rm T}$. 
Since  by setting $\Delta=0$
we have neglected dissipation, the diagonal
elements of the reduced density matrix remain unchanged, that is, one has
\begin{eqnarray}
\rho_{++}(t)&=&|\alpha|^2|N_+(t)|^2\int \mathcal{D}\sigma  \exp\left[-\int
  d^3p\ \sigma(\mathbf{p})\Re \Omega_+(t)\sigma(-\mathbf{p})\right]\nonumber\\
&=&|\alpha|^2=\rho_{++}(0)
\end{eqnarray}
and analogously $\rho_{--}(t)=\rho_{--}(0)$.
The probabilities of finding the
system in state $+$ or $-$ are thus unchanged by the environment; this
corresponds to a quantum-nondemolition (or ideal) measurement
\cite{deco,schlosshauer}.  

The non-diagonal elements can be calculated as follows:
\begin{eqnarray}
\rho_{+-}(t)&=&\rho_{-+}^*(t)=
\alpha \beta^*N_+ \left(N_-\right)^*\times \nonumber\\ \;\; & & 
\int D\sigma \exp\left(-\frac{1}{2}\int d^3
  p\ \sigma(\mathbf{p})\left(\Omega_++\left(\Omega_-\right)^*\right)
\sigma(-\mathbf{p})-i(E_+-E_-)  
  t\right)\nonumber\\   
&=&\alpha\beta^*\frac{\det^{1/4}(\Re\Omega_+)\det^{1/4}(\Re\Omega_-)}{\det^{1/2} 
\left(\left(\Omega_++\left(\Omega_-\right)^*\right)/2\right)}\times\nonumber\\  
& &\times\exp\left(-i\int^t dt' 
  \frac{1}{2a^3}\mathrm{Tr}(\Re\Omega_+-\Re\Omega_-)-i(E_+ 
  -E_-) t\right)\,.  
\end{eqnarray}
 From \eqref{scaleequ2} we can see that 
the functions $\Omega_{+/-}$ depend on the small parameter 
$a^2\gamma/p^2$ for sub-Hubble modes and on $\gamma/H^2$
for super-Hubble modes, see below for an explanation of these terms.
Expanding $\Omega_{+/-}$ up to second order in $\gamma$ yields
\begin{eqnarray}
\Omega_{+/-}(\gamma)&\approx&\Omega_{+/-}\Bigg|_{\gamma=0}+\frac{d}{d
  \gamma}\Omega_{+/-}\Bigg|_{\gamma=0}\gamma+\frac{1}{2}\frac{d^2}{d
  \gamma^2}\Omega_{+/-}\Bigg|_{\gamma=0}\gamma^2\nonumber\\ 
&=:&\Omega+\Omega_{\gamma}\gamma+\frac{1}{2}\Omega_{\gamma\gamma}\gamma^2\ ,
\end{eqnarray}
where derivatives with respect to $\gamma$ are denoted by indices.
(Strictly speaking, the expansion is with respect to the dimensionless
combinations $a^2\gamma/p^2$ or $\gamma/H^2$.)  

The approximate expression for the non-diagonal elements then reads
\begin{eqnarray}\label{nondiag}
\rho_{+-}(t)=\rho_{+-}(0)\exp\left(-\frac{g_{\rm
      s}^2(\phi_+-\phi_-)^2}{4}
\mathrm{Tr}\left(\frac{(\Re\Omega_{\gamma})^2
+(\Im\Omega_{\gamma})^2}{(\Re\Omega)^2}\right)-i\varphi_{+-}\right)       
\end{eqnarray}
with $\rho_{+-}(0)=\alpha\beta^*$ and
\begin{eqnarray}
 \varphi_{+-}&=&\mathrm{Tr}\left(\frac{\Im\Omega_{\gamma}}{\Re\Omega}\frac{g_{\rm
       s}(\phi_+-\phi_-)}{2}+  
\left(\frac{\Im\Omega_{\gamma\gamma}}{\Re\Omega^S}
-\frac{\Im\Omega_{\gamma}\Re\Omega_{\gamma}}
{(\Re\Omega)^2}\right)\frac{g_{\rm  
    s}^2(\phi_+^2-\phi_-^2)}{2}\right)\nonumber\\  
&+&\int^t d\tilde{t}\ 
\frac{1}{2a^3}\mathrm{Tr}(\Re\Omega_+-\Re\Omega_-)+(E_+
-E_-) \tilde{t}\,. 
\end{eqnarray}

In the following, we want to discuss the explicit form of the 
decoherence factor.
To begin with, we consider the impact of the {\em sub-Hubble modes}
on the system, that is, the impact of modes whose wavelength
$\lambda=2\pi a/p$ is smaller than the Hubble scale $H^{-1}$. 
Using a WKB-approximation, which is adequate for $p^2\gg a''/a$,
the solutions to the differential equation (\ref{scaleequ2}) read
\begin{eqnarray}
u_{+/-}(p,\eta)=\frac{A}{a}\exp\left(i\int^\eta
  d\tilde{\eta}\ \omega_{+/-}(\tilde{\eta})\right) \ ,
\end{eqnarray}
where
\begin{eqnarray}
\omega_{+/-}(\eta):=\left(p^2+\gamma a^2-\frac{a''}{a}\right)^{1/2}\,. 
\end{eqnarray}
We have chosen an adiabatic vacuum in the infinite past for each Fourier mode
${\bf p}$. This choice is possible because the interaction vanishes for 
$a\rightarrow 0$ when the modes are far inside the horizon. 
In the case of a massive scalar field ($g_s\phi_{+/-} > 0$) in the
exact de Sitter background, it leads to the Bunch--Davies vacuum for the whole 
field $\sigma$, see e.g. \cite{Parker}.
The trace of the real part of the exponent in (\ref{nondiag}) is
\begin{eqnarray}\label{subhub}
& &\mathrm{Tr}\left(\frac{(\Re\Omega_{\gamma})^2
+(\Im\Omega_{\gamma})^2}{(\Re\Omega)^2}\right)=\frac{\pi     
  a^4V}{(2\pi)^3}\int_{p_{\mathrm{min}}}^\infty
dp\ \frac{p^2}{\left(p^2-\frac{a''}{a}\right)^2}\nonumber\\   
&=&\frac{ 
  a^4V}{16\pi^2}\left(\frac{p_{\mathrm{min}}}{p_{\mathrm{min}}^2-\frac{a''}{a}}
+\frac{1}{2}\sqrt{\frac{a}{a''}} 
\ln\left(\frac{p_\mathrm{min}+\sqrt{\frac{a''}{a}}}{p_\mathrm{min}
-\sqrt{\frac{a''}{a}}}\right)\right)\, ,    
\end{eqnarray} 
where $V$ denotes the dimensionless coordinate volume which must be
fixed by an appropriate infrared cutoff.
The WKB-approximation holds for modes far inside the horizon,
$p_\mathrm{min}>\sqrt{a''/a}$; 
in the case of a constant Hubble rate, we have
$p_\mathrm{min}>\sqrt{2}H a$.

To discuss the explicit form of the decoherence rate for {\em super-Hubble 
modes}, that is, for modes with wavelengths greater than the Hubble
scale, we shall 
restrict ourselves to the expanding de~Sitter case where $H=\mathrm{constant}.$ 
In this case, it is not possible to find WKB solutions valid
for all times including $t\to\infty$ ($\eta\to 0)$, because the time
evolution for super-Hubble modes 
is highly nonadiabatic. The solution of Eq.~(\ref{scaleequ2}) that has
the correct WKB behavior for $\eta\to -\infty$ is then given by
\begin{eqnarray}\label{sol}
u_{+/-}(p,\eta)=\frac{H\sqrt{\pi}}{2}\left(\frac{\eta}{p}\right)^{3/2}
\mathcal{H}^{(1)}_{\sqrt{9/4-\gamma/H^2}}(p\eta)
\ ,
\end{eqnarray}
where $\mathcal{H}^{(1)}$ denotes a Hankel function. (Recall that $\gamma<H^2$.)
In the massless case $g_{\rm s}=0$, this expression reduces to the textbook
free field solution for the adiabatic (WKB) initial vacuum state at $\eta\to -
\infty$:
\begin{eqnarray}\label{Bunch}
u_{g_{\rm s}=0}(p,\eta)=
-\frac{H\eta}{\sqrt{2}p^2}\left(1+\frac{i}{p\eta}\right)e^{ip\eta}\,.   
\end{eqnarray}
Note that due to the absence of WKB solutions for $t \to \infty$ ($\eta\to 0$),
there exists an ambiguity in the definition of a particle for super-Hubble modes
for sufficiently small masses $m^2=\gamma^2\le 9H^2/4$. In
addition, these  
modes are strongly squeezed. In particular, in the massless case
$g_s=0$ we have
\begin{eqnarray}\label{particle}
\sinh^2r_p=\frac{a^2 H^2}{4 p^2}\, ,
\end{eqnarray}
where $r_p$ is the squeezing parameter of the mode, see e.g. \cite{PS96,KP09}.
On the other hand, in the exact de Sitter case it is possible, in principle,
to consider the whole solution (\ref{Bunch}) as the second-order corrected WKB
solution that would imply no massless ``particle creation'' in the de Sitter 
background, see e.g. \cite{AMM05}. However, any ambiguity disappears if we take 
into account that $H$ is not constant in any realistic inflationary model and 
decreases after the end of inflation (or even during it), see the more extended
discussion of this problem in \cite{KLPS07}.   
 
Using (\ref{particle}), we can express $\Omega(p,\eta)$ 
entirely as a function of the squeezing parameter:
\begin{eqnarray}\label{omega}
\Omega(p,t)=\frac{p a^2}{1-2 i\sinh r_p}\,.
\end{eqnarray}
Using (\ref{omega}) and the approximation of (\ref{sol}) for
$p\eta\rightarrow0$, 
\begin{eqnarray}
u_{+/-}(p\eta)\propto
\left(\frac{\eta}{p}\right)^{3/2}
\left(\frac{p\eta}{2}\right)^{-\sqrt{9/4-\gamma/H^2}}\,, 
\end{eqnarray}
we obtain for the impact of super-Hubble modes on the system the result
\begin{eqnarray}\label{superhubble} & & 
\mathrm{Tr}\left(\frac{(\Re\Omega_{\gamma})^2+
(\Im\Omega_{\gamma})^2}{(\Re\Omega)^2}\right)=   
\frac{
  V a^2}{18\pi^2 H^2}\int_{p_\mathrm{min}}^{p_\mathrm{max}}
dp\ (1+4\sinh^2r_p)^2\nonumber\\ 
& &\; =\frac{V a^2}{18\pi^2
  H^2}\left[p_\mathrm{max}-p_\mathrm{min}-\right.\nonumber\\ & & \;\; \left.
\left(\frac{2a^2H^2}{p_\mathrm{max}}-\frac{2a^2H^2}{p_\mathrm{min}}\right)
-\left(\frac{a^4H^4}{3p^3_\mathrm{max}}-
\frac{a^4H^4}{3p^3_\mathrm{min}}\right)\right]\,. 
\end{eqnarray}
This result for the trace depends on the minimal and maximal values
for the dimensionless wave number. For the super-Hubble modes, we take
for the minimal wavelength the Hubble scale, so $p_\mathrm{max}=2\pi
aH$. What has to be taken for the maximal wavelength? 
Clearly, an infinite wavelength would lead to $p_\mathrm{min}=0$ and
thus to a divergence in \eqref{superhubble}. In a closed universe, a
reasonable cutoff would be $\lambda=a$. In the case of an open
universe, we shall adopt the argument on p.~159 of \cite{Linde}, which
goes as follows. We assume that the initial fluctuation spectrum
has a cutoff at $\lambda_0\sim H^{-1}$, where the initial size of the
inflating region is about $H^{-1}$. This leads to a cutoff
$\lambda=a$ as in the closed case and we can set $p_\mathrm{min}=2\pi$
in \eqref{superhubble}.

However, in the case of a real post-inflationary universe, it is
natural to take for the cutoff of $\lambda$ its {\em homogeneity
  scale}, that is, the scale, at which perturbations generated during
inflation become of order unity and the spacetime description using
the FRW background \eqref{FRW} loses sense. Though this scale is much
smaller than the radius of pre-inflationary curvature, it is still
much (typically exponentially) larger than the Hubble scale after the
end of inflation.

Evaluating the trace using these numbers and inserting the result into 
(\ref{nondiag}), we find for the absolute value of the non-diagonal element 
of the density matrix:
\be
\lb{rhosuper}
\vert\rho_{+-}(t)\vert = \vert\rho_{+-}(0)\vert
\exp\left(-\frac{g_{\rm
      s}^2a^2V(\phi_+-\phi_-)^2}{72\pi^2H^2}\left[\frac{(aH)^4}{24\pi^3} 
    +{\mathcal O}(aH)^2\right]\right).
\ee
We recognize explicitly that this non-diagonal element becomes
increasingly small for increasing $aH$, that is, decoherence becomes
efficient and the field is localized in one or the other well.

Evaluating in the same manner the density matrix (\ref{subhub}) for
the sub-Hubble modes, we obtain instead
\be
\lb{rhosub}
\vert\rho_{+-}(t)\vert = \vert\rho_{+-}(0)\vert
\exp\left(-\frac{g_{\rm s}^2a^3VC(\phi_+-\phi_-)^2}{64\pi^2H}\right)\ ,
\ee
where $C\approx 0.329575$. Taking the ratio of
the widths of the two Gaussians 
(\ref{rhosuper}) and (\ref{rhosub}), we get
\be
\frac{(\Delta\phi)^2_{\rm super}}{(\Delta\phi)^2_{\rm sub}}\approx
\frac{8.9\pi^3}{(aH)^3} \ ,
\ee
which goes to zero for $aH\to\infty$, that is, the super-Hubble modes
are much more efficient in the localization of $\phi$ than the
sub-Hubble modes. The reason for this is that the ensuing entanglement
with the DE field is stronger with the super-Hubble modes due to the
stronger interaction. 

One can associate with \eqref{rhosuper} a decoherence time 
roughly as follows. Assuming that $a(t)=H^{-1}\exp(Ht)$, the condition
that the (absolute value of the) exponent becomes of order unity leads
to a decoherence time $t_{\rm d}\sim (6H)^{-1}$ times logarithmic terms
containing the coupling and the separation of the minima
(counted from the beginning of the de~Sitter stage).

The environmental field $\sigma$ used in the investigation above can
have different origins. The most natural one is given by the 
scalar and tensorial modes of the metric itself. The dynamics of these
modes during an exponential expansion of the background is well
known and can be described, for example, in the squeezed-state
formalism already mentioned above \cite{PS96}. If the dynamics of
these modes is unitary, they become highly squeezed in the field momentum
and highly elongated in the field variable. However, these modes
are themselves prone to decoherence \cite{KPS98,KP09,KLPS07}. The
classical pointer basis distinguished by the interaction of these
modes to their environment is the field-amplitude basis; this is why
one can describe the primordial fluctuations by classical stochastic
field variables. Why, then, can these fluctuations serve as a quantum
environment to decohere the DE field?

Let us assume that the modes $\sigma_p$ couple to another field with
modes $\chi_k$ responsible for their decoherence. Unless the coupling
is very strong (which would not be realistic), the total quantum state
is of the form
\be
\lb{totalstate}
\Psi(\phi,\{\sigma_p\},\{\chi_k\})
\approx \psi_1(\phi,\{\sigma_p\})\psi_2(\{\sigma_p\},\{\chi_k\})\ .
\ee
The entanglement in the wave function
$\psi_2(\{\sigma_p\},\{\chi_k\})$ is responsible for the decoherence
of the modes $\sigma_p$ \cite{KPS98}. However, tracing out the modes
$\chi_k$ in the full state \eqref{totalstate} is ineffective as far as
the DE field $\phi$ is concerned; for the decoherence of the latter
only the wave function $\psi_1(\phi,\{\sigma_p\})$ is
responsible. That is, our results presented in this section are
insensitive of the modes $\sigma_p$ being themselves decohered or not
(provided, of course, the involved interactions are not too strong).

This is in accordance with decoherence in quantum mechanics
\cite{deco}. Scattering processes can localize a quantum system such
as an electron or a molecule. The important point is that the
scattering causes entanglement between the scattered system and the
scattering agency and that the information about the superposition is
no longer available at the system itself. Typically, the environment
is itself decohered, but since the full states are usually of the form
\eqref{totalstate}, this decoherence is irrelevant for the decoherence
of the original system. A somewhat related example is the one
discussed in \cite{bath}. There, two oscillators become entangled with
the same heat bath. However, unless the two oscillators are close
together, the respective entanglements with the bath will not lead to
an entanglement between the oscillators themselves, that is, the total
state will be of a form similar to \eqref{totalstate}.

To summarize, we have shown in this section that it is justified to
assume that the DE field is localized in one of the two
wells only; interference terms between the two wells are dynamically
suppressed by decoherence. This justifies the scenario presented by
Yokoyama that the small positive cosmological constant could arise
from the DE field not being in its ground state -- it is localized by
decoherence in one of the minima of the potential.

An extension of Yokoyama's work to the case of many wells, taking into 
account ideas from string theory, was suggested in \cite{KPZ}. 
There, the authors assume the ground state of our Universe
to be a superposition of all accessible vacuum states.
However, this seems to be a doubtful assumption, since
the unavoidable interaction with
environmental degrees of freedom, for example
Standard Model fields or thermal excitations, should lead to their
decoherence. 

\section{Tunneling rate}\label{tunnel}

In this section we shall investigate the influence of the
system--environment interaction on the tunneling rate.
Before we discuss this in detail, we want to recall some basic facts about
tunneling in field theory. We shall first address the case of a
Minkowski background and then turn to the expanding Universe.

Following \cite{coleman}, we know that the tunneling rate of a system
given by a scalar field Lagrangian is of the form
\begin{eqnarray}
\lb{tunnelrate}
\Gamma_0=A\exp(-S_E(\phi))\ ,
\end{eqnarray}
where $S_E(\phi)$ is the classical Euclidean action of the scalar field
evaluated along the tunneling trajectory of $\phi$; the prefactor $A$
can be determined by the second variation of the action.

According to the classical equations of motion, the field $\phi$
adopts for most of the time  
the value $\phi_+$ of the false vacuum and approaches
the value $\phi_-$ of the true vacuum after a short transition time.
The terms ``false vacuum'' and ``true vacuum'' may be misleading, since
they denote the classical minima of the potential (with $\phi_+$
representing here the higher minimum), in contrast to the true
quantum mechanical vacuum which is a superposition of $\phi_+$ and
$\phi_-$.

The tunneling time $T$ between the two vacuum states is assumed to be
large compared to the characteristic instanton transition time
$1/\omega$. We thus consider situations in which various
tunneling processes from one minimum to the other can be considered separately. 
This picture of separated transitions can only be justified 
by decoherence, since the superposition principle is universally valid
and thus holds also for widely separated ``jumps''.

Assuming spherical symmetry, a transition between the localized vacuum states 
can be described by the growth of a bubble that is nucleating
spontaneously at a radius $R_0$. 
The true vacuum inside and the false vacuum outside the vacuum bubble
are separated by 
a wall with a negative surface tension.
In the limit of a small energy difference between the localized vacuum states,
the nucleation radius is given in terms of the energy difference and
the surface tension \cite{coleman}. 

The interaction with the environment will influence the decay rate 
due to decoherence and dissipation.
The authors of \cite{caldlegg} considered a macroscopic position variable 
 coupled to a heat bath of harmonic oscillators. 
After integrating out the environmental degrees of freedom, they
obtained a Langevin equation with an effective friction term resulting 
from dissipative effects.
The consequence of this friction term is to {\em reduce} the tunneling
amplitude. 

In contrast to \cite{caldlegg} we consider the tri-linear interaction
(\ref{trilinear}) between system and environment.
In addition we work in the context of field theory and thus
 do not restrict ourselves to a quantum mechanical model, that is, to
a finite number of oscillators.

Starting from (\ref{22}) and switching to Euclidean time, $t\rightarrow-i t_E$,
we find the Euclidean action 
\begin{eqnarray}
S_E=S_{E,{\rm sys}}+\int dx^3 dt_E\, \sigma(\mathbf{x},t)\left(-\frac{1}{2}
  \Box_E+g_{\rm s} \phi(t_E)\right)\sigma(\mathbf{x},t) \,. 
\end{eqnarray}
Integrating out the environmental field $\sigma$ leads to the
following formal expression that modifies the tunneling rate
\eqref{tunnelrate}, 
\begin{eqnarray}\label{31}
\Gamma=\Gamma_0N\times\int D\sigma(\mathbf{x},t)
\exp(-S_E)=\sqrt{\frac{\mathrm{Det}(-\Box_E)}{\mathrm{Det}(-\Box_E+2
    g_{\rm s} \phi(t_E))}}\Gamma_0\ .
\end{eqnarray}
The normalization $N$ was chosen such that $\Gamma|_{g_{\rm s}=0}=\Gamma_0$.
Solving the nonlocal equations of motion given by the variation of
\begin{eqnarray}
S_{E,\mathrm{eff}}=S_{E,{\rm sys}}+\frac{1}{2}\mathrm{Tr}\ln(-\Box_E+2 g_{\rm s} \phi)
\end{eqnarray}
and evaluating the effective action along the tunneling trajectory of $\phi$
would give the exact modified tunneling amplitude.
In order to simplify the calculations, we want to assume that at lowest order 
the trajectory of $\phi(t_E)$ is given by the unperturbed equations of motion.
In this approximation we can compute the functional determinant.

For this computation we have to 
solve the eigenvalue equation 
\begin{eqnarray}
\left(-\frac{d^2}{dt_E^2}-\triangle +2g_s\phi(t_E)\right)\psi=\lambda\psi\,.
\end{eqnarray}
Since the nucleation of the bubble is spherically symmetric, it is appropriate
to use the Laplacian in spherical coordinates.
Making the ansatz $\psi=\phi_{ln}(r)Y_{lm}(\theta,\phi)u(t_E)$, we find for the
eigenvalue equation of the radial component
\begin{eqnarray}\label{spherical}
\left(-\frac{1}{r^2}\frac{\partial}{\partial r}r^2\frac{\partial}{\partial r}+
\frac{l(l+1)}{r^2}\right)\phi_{ln}(r)=\kappa_{nm}\phi_{ln}(r)\,.
\end{eqnarray}
A natural boundary condition would be $\phi_{ln}(R_0)=0$, that is, the
eigenfunctions are vanishing at the boundary of the bubble. 
Equation (\ref{spherical}) is solved by the spherical Bessel
functions, that is, $\phi_{ln}(r)\propto j_l(\kappa_{ln}r)$. 
The eigenvalues $\kappa_{ln}$ are the $n$-th root of $j_l$ divided by $R_0$.
We thus have
\begin{eqnarray}
\mathrm{Det}(-\Box_E+2 g_s
\phi(t_E))=\prod_{n,l=0}^{\infty}(2l+1)
\mathrm{Det}_{nl}\left(-\frac{d^2}{dt^2_E}+\kappa_{ln}^2+2 
  g_s\phi(t_E)\right)\,, 
\end{eqnarray}
where the degeneracy of the eigenvalues was taken into account with a
factor $2l+1$. 

Since the eigenvalues of the spherical Bessel functions are not
explicitly known, we simplify the problem by assuming periodic  
boundary conditions in a volume $L^3$ with $L=\mathcal{O}(R_0)$.
For the spatial part of the Euclidean d'Alembert operator, we choose
periodic boundary conditions with a length $L$.
The functional determinant involved in (\ref{31}) then separates into
a product of 
functional determinants labeled by the mode number $n$:
\begin{eqnarray}\label{30}
\mathrm{Det}(-\Box_E+2 g_{\rm s}
\phi(t_E))=\prod_{n=0}^\infty 4\pi
n^2\mathrm{Det}_n\left(-\frac{d^2}{dt_E^2}+\frac{4\pi^2 
    n^2}{L^2}+2 g_{\rm s}\phi(t_E)\right)\,. 
\end{eqnarray}
This expression is divergent for two reasons. First, for
each fixed mode number $n$, the determinant is an infinite product of
eigenvalues, which is in general infinite.
Second, due to infinitely many modes, the situation becomes even
worse.

In order to regularize the expression (\ref{30}) we choose the $\zeta$-function
regularization method presented in \cite{barkamkar}.
A short summary of this method is given in the Appendix.
For any second-order differential operator $\mathcal{D}$ we can write 
\begin{eqnarray}\label{100}
(\mathrm{Det}\mathcal{D})^{1/2}=
\exp\left(-\frac{1}{2}\zeta'(0)-\frac{1}{2}\zeta(0)\ln
  \mu^2\right)\ ,  
\end{eqnarray}
where $\zeta(s)$ is the generalized zeta function
\begin{eqnarray}\label{101}
\zeta(s)=\sum_\lambda  \frac{1}{\lambda^s}
\end{eqnarray}
involving all eigenvalues $\lambda$ of the differential operator. 
The parameter $\mu$ appearing in (\ref{100}) is a
renormalization parameter with dimension of a mass.
According to \cite{barkamkar}, the $\zeta$-function reads
\begin{eqnarray}\label{zetaint}
\zeta(s)=\frac{\sin(\pi s)}{\pi}\int_0^\infty \frac{d
  M^2}{M^{2s}}\frac{d^2}{dM^2}I(M^2,s)
\end{eqnarray}
with
\begin{eqnarray}
I(M^2,s)=\sum_{n=1}^{\infty} \frac{1}{n^{2s}}\ln u(M^2 n^2,t_E)\,.
\end{eqnarray}
The integral (\ref{zetaint}) converges for some $s>0$, since
$I(M^2,s)$ increases with finite  
polynomial order proportional to $M^k$, and can be analytically
continued to $s=0$ \cite{barkamkar}. 

The functions $u(-\lambda,t_E)$ are the eigenfunctions of the
differential operator under consideration. 
The eigenvalue equation corresponding to (\ref{30}) reads
\begin{eqnarray}\label{102}
\left(-\frac{d^2}{dt_E^2}+\frac{4\pi^2 n^2}{L^2}+2 g_{\rm
    s}\phi(t_E)\right)u(-\lambda,t_E)=\lambda\, u(-\lambda,t_E)\,. 
\end{eqnarray}
Note that the differential operator is always positive definite since
$g_{\rm s}\phi(t_E)>~0$. 

In general one needs two independent boundary conditions in order to
determine the eigenvalues of (\ref{102}) 
uniquely. We assume that the mode functions have a root at nucleation
time $T_0$, that is, $u(-\lambda,T_0)=0$ \cite{barkamkar}. 
This time is usually of the order of or equal to the nucleation radius
$R_0$ \cite{coleman}. 
The second boundary condition is given by the normalization, see below.
Since the regularization method employed discards one of the two
independent solutions of equation (\ref{102}) (see the Appendix for a
detailed discussion), the eigenfunctions are uniquely determined by a
normalization condition.
 
The leading term of the uniform WKB-solution of (\ref{102}) reads
\begin{eqnarray}
u(-\lambda,t_E)&=&\left(\frac{4\pi^2 n^2}{L^2}+2 g_{\rm
    s}\phi(t_E)-\lambda\right)^{-1/4}\nonumber\\ 
& &\times\exp\left[\int^{t_E}_{t_{E,0}} dt'_E\left(\frac{4\pi^2
      n^2}{L^2}+2 g_{\rm s}\phi(t'_E)-\lambda\right)^{1/2}\right] ,
\end{eqnarray}
where the choice of $t_{E,0}$ determines the normalization of $u(-\lambda,t_E)$.
We assume that the scalar field $\phi$ is in the true vacuum at some
large positive time $t_E$, that is, $\phi(t_E)=\phi_-$, and fix the
normalization such that 
\begin{eqnarray}\label{107}
u(-\lambda,t_E)&=&\left(\frac{4\pi^2 n^2}{L^2}+2 g_{\rm
    s}\phi_--\lambda\right)^{-1/4}\nonumber\\ 
& &\times\exp\left[-\left(\frac{4\pi^2 n^2}{L^2}+2 g_{\rm
      s}\phi_--\lambda\right)^{1/2}t_E\right]\,. 
\end{eqnarray}
Since the normalization of $u(-\lambda,t_E)$ enters the function
$I(M^2,s)$ only logarithmically, 
the choice of a different normalization will not significantly change the
results. 

For large mode numbers $n$ there is an approximate degeneracy of $4\pi n^2$.
We find
\begin{eqnarray}\label{infinitesum}
I(M^2,s)&=&\sum_{n=1}^\infty \frac{4\pi
  n^2}{n^{2s}}\Bigg[-\frac{1}{4}\ln\left(\frac{4\pi^2n^2}{L^2}+2g_{\rm
    s}\phi(t_E)+M^2n^2\right)\nonumber\\ 
& &+\left(\frac{4\pi^2 n^2}{L^2}+2 g_{\rm
    s}\phi(t_E)+M^2 n^2\right)^{1/2}t_E\Bigg]\,. 
\end{eqnarray}
For $M^2\rightarrow0$ we define 
\begin{eqnarray}
I(0,s)=:\sum_{n=1}^\infty\frac{f(n)}{n^{2s}}
\end{eqnarray}
with
\begin{eqnarray}
f(n)=-2\pi n^2\ln n+g(n)
\end{eqnarray}
and
\begin{eqnarray}
g(n)&=&-\pi n^2\ln\left(\frac{4\pi^2}{L^2}+\frac{2 g_{\rm
      s}\phi(t_E)}{n^2}\right)\nonumber\\ & & +4\pi
n^2\left(\frac{4\pi^2 n^2}{L^2}+2g_{\rm
    s}\phi(t_E)\right)^{1/2}t_E\,. 
\end{eqnarray}
The function $I(0,s)$ can be evaluated using the Abel--Plana formula
\cite{AbelPlana,barkamkar}
\begin{eqnarray}\label{103}
I(0,s)&=&-\sum_{n=1}^\infty\frac{2\pi n^2\ln n}{n^{2s}}+\int_0^1 dn\,
g(n)+\int_1^\infty d n\,\frac{g(n)}{n^{2s}}\nonumber\\ 
& &+i\int_0^\infty dy\frac{g(i y)-g(-iy)}{e^{2\pi
    y}-1}-\frac{1}{2}g(0)+\mathcal{O}(s)\,. 
\end{eqnarray}
The sum on the right-hand side of (\ref{103}) will not affect the
$\zeta$-function, since it corresponds to   
an $M$-independent term of $I(M^2,s)$.
We have retained in (\ref{103}) the regularizing factor $1/n^{2s}$
only in the sum and in the second integral, since the remaining terms are  
finite for $s\rightarrow 0$.
The splitting of the integral at $x=1$ is made for convenience and
does not affect the final result.
In order to regularize the remaining integral, 
we have to integrate by parts several times.
For this purpose we change the integration variable to $x=1/n$ and
define the function 
\begin{eqnarray}
 F(x)=x^3 g(1/x)\ ,
\end{eqnarray}
which is analytic at $x=0$.
The function $I(0,s)$ can be expanded according to
\begin{eqnarray}
I(0,s)=\frac{I^\mathrm{pole}(0)}{s}+I^\mathrm{R}(0)+\mathcal{O}(s)\,.
\end{eqnarray}
Performing integrations by parts and discarding the boundary terms
where we cannot interchange the limits
$x\rightarrow0$ and $s\rightarrow0$ \cite{barkamkar}, we find
\begin{eqnarray}\label{104}
I^\mathrm{pole}(0)=\frac{1}{48}\frac{d^4 F(x)}{dx^4}\Bigg|_{x=0}
\end{eqnarray}
and
\begin{eqnarray}\label{105}
I^\mathrm{R}(0)&=&\frac{25}{288}\frac{d^4
  F(x)}{dx^4}\Bigg|_{x=0}-\frac{1}{24}\int_0^\infty dx \ln x 
\frac{d^5 F(x)}{dx^5}\nonumber\\
& &+i\int_0^\infty dy\frac{g(i y)-g(-iy)}{e^{2\pi y}-1}-\frac{1}{2}g(0)-\sum_{n=1}^\infty\frac{2\pi n^2\ln n}{n^{2s}}\,.
\end{eqnarray}
The expression (\ref{104}) and the second integral in (\ref{105})
result from the convergent integral that remains  
after the integrations by parts.
A factor $x^{2s}/s$ in this integral can be expanded according to
$x^{2s}/s=1/s+2\ln x+\mathcal{O}(s)$ which leads to the aforementioned 
terms. 

The first term in (\ref{105}) results from the finite contributions of
the boundary terms where the limit $x\rightarrow 0$ 
and $s\rightarrow 0$ can be interchanged.
We find the explicit expressions
\begin{eqnarray}
I^\mathrm{pole}(0)=- \frac{g_{\rm s}^2L^3t_E \phi^2(t_E)}{8\pi^2}\,,
\end{eqnarray}
and
\begin{eqnarray}\label{106}
I^R(0)&=&\frac{g_{\rm s} L^3 \phi(t_E)}{24\pi^2}\bigg[2\sqrt{2\pi
  g_{\rm s} \phi(t_E)}\\ 
&+&g_{\rm s}\phi(t_E)t_E\left(14+3\ln\left(\frac{L^2g_{\rm
        s}\phi(t_E)}{2^3\pi^2}\right)\right)\bigg]-\frac{25}{48}\frac{g_{\rm
    s}^2L^3\phi^2(t_E)t_E}{\pi^2}\nonumber\\ 
&-&2 \Re \left\{i \pi \int_0^{\frac{L}{\pi}\sqrt{\frac{g_{\rm
          s}\phi(t_E)}{2}}}\frac{y^2}{e^{2\pi
      y}-1}\ln\left(1-\frac{L^2 g_{\rm s}\phi(t_E)}{2\pi^2
      y^2}\right)\right\}\nonumber\\ 
&+&8\pi \int_{\frac{L}{\pi}\sqrt{\frac{g_{\rm s}\phi(t_E)}{2}}}^\infty
dy\frac{y^2\left(\frac{4\pi^2 y^2}{L^2}-2 g_{\rm
      s}\phi(t_E)\right)^{1/2}t_E}{e^{2\pi y}-1}-\sum_{n=1}^\infty
\frac{2\pi n^2 \ln n}{n^{2s}}\nonumber\,. 
\end{eqnarray}
Furthermore, we need the pole part of $I(M^2,0)$ for $M\neq 0$.
Equation (\ref{104}) also holds for arbitrary $M$ if we replace
$4\pi^2/L^2\rightarrow4\pi^2/L^2+M^2$ in the definition 
of $g(n)$. 
More concretely, we may expand $I(M^2,s)$ for large $n$ and use the
fact that the Riemann $\zeta_R$-function has 
a pole at $s=1$, 
\begin{eqnarray}
\zeta_R(2s+1)=\frac{1}{2s}+\mathcal{O}(s^0)\,.
\end{eqnarray}
We find
\begin{eqnarray}\label{111}
I^\mathrm{pole}(M^2)=-\frac{g_{\rm s}^2\pi t_E
  \phi^2(t_E)}{\left(\frac{4\pi^2}{L^2}+M^2\right)^{3/2}}\,. 
\end{eqnarray}
In the limit $M\rightarrow\infty$ we find the regular part of $I_R(M^2,0)$ to be
\begin{eqnarray}
I^\mathrm{R}(M^2\rightarrow\infty)=-\sum_{n=1}^\infty\frac{2\pi n^2\ln n}{n^{2s}}\,.
\end{eqnarray}
This sum appears also in $I(0,s)$ and will therefore not affect
$\zeta(0)$ and $\zeta'(0)$. 
The logarithmic contribution of $I(M^2,0)$ vanishes, since
$\zeta_R(-2)=0$.

Using (\ref{100}) and Eq. (\ref{A6}) from the Appendix, we can now compute
the modified tunneling amplitude from $I(M^2,s)$. 
Expanding (\ref{106}) up to the first order in $g_{\rm s}$ and using
the normalization defined through (\ref{107}), we find the modified
tunneling amplitude 
\begin{eqnarray}
\lb{modifiedtunneling}
\Gamma=\Gamma_0\exp\left(-\frac{g_{\rm
      s}\phi(t_E)L^2}{8\pi}+\frac{g_{\rm s} L\phi(t_E)t_E}{12}
  \right)\,. 
\end{eqnarray}
Let us illustrate this result by using a definite expression for the
function $\phi(t_E)$. With the choice
\begin{eqnarray}
\phi(t_E)=\frac{\phi_--\phi_+}{2}\tanh(\omega t_E)+\frac{\phi_-+\phi_+}{2},
\end{eqnarray}
where $1/\omega$ is the characteristic time of the instanton,
\eqref{modifiedtunneling} becomes 
in the limit of large Euclidean nucleation time $T_0$, that is,
$T_0\gg 1/\omega$, 
\begin{eqnarray}\label{108}
\Gamma\approx\Gamma_0\exp\left(-\frac{g_{\rm s}\phi_-
    L^2}{8\pi}+\frac{g_{\rm s}\phi_- LT_0}{12}\right)\, .
\end{eqnarray}
The first term in the exponent, which results from
subexponential terms of the WKB-expansion, is negligible in the limit
$T_0\gg L$. 

The result (\ref{108}) deserves some explanation.
The appearance of the quantization length $L$ is due to the fact that
the environmental field enters the interaction quadratically.
Integrating over all momenta leads to an expression which increases
with the quantization volume.
If we had chosen a bilinear coupling, the end result would have been
independent of the quantization 
volume, since the environmental field would fluctuate around zero.

If we neglect the first term in the exponent of (\ref{108}), the
tunneling amplitude will always be enhanced, since the product $g_{\rm
s}\phi$ is positive;
a negative sign would render the model unstable.

Recalling the results of \cite{caldlegg},
 one would expect a reduction of the tunneling rate due to dissipative
effects. In \cite{caldlegg} the interaction
with the environment leads to an effective 
friction term in the equation of motion for the system variable.
Since we consider an interaction that is quadratic in the
environmental field, the situation here, however, is different.\footnote{In 
this connection, note also the recent paper \cite{BGR10} where
an enhancement of the tunneling rate was found for the same mathematical
reason, though in a different physical situation, with the electromagnetic 
field instead of an environmental one.}

A finite number of environmental degrees of freedom ($N$ harmonic oscillators)
 would lead to a suppression of the tunneling rate, that is, to a
 negative sign in the second term of the exponential in (\ref{108}). 
The mathematical reason is that for $N$ oscillators we have in the
leading term the sum 
\begin{eqnarray}
\sum_{n=1}^N n=\frac{N}{2}(N+1)\ ,
\end{eqnarray}
which is always positive. In contrast,
an infinite number of environmental degrees of freedom leads to a sum
that must be regularized, e.g. with the help of the 
Riemann $\zeta_R$-function as above. One then gets
\begin{eqnarray}
\sum_{n=1}^\infty \frac{n}{n^{2s}}=-\frac{1}{12} \quad\text{for}\quad
s\rightarrow 0 \ ,
\end{eqnarray}
that leads to the positive sign in (\ref{108}).
Physically this can be interpreted analogously to the Casimir effect: due
to the boundary conditions defining the functional determinant, there
are fewer environmental modes than there would be without the
boundary conditions, and thus there is less decoherence. This is a
result that one would not expect on purely quantum mechanical (as
opposed to field theoretical) calculations. 

\begin{figure}
\centering
\includegraphics{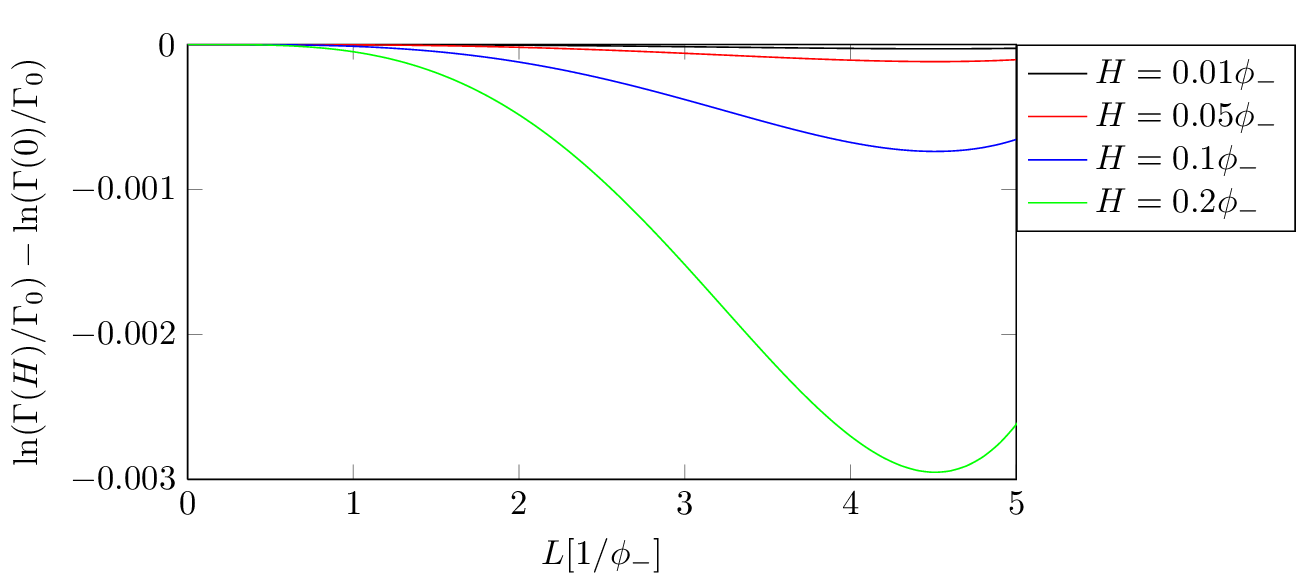}
\caption{We set $L=T_0$ for simplicity and show the change of the
  modified tunneling rate 
  with increasing $L$ for different Hubble parameters $H$. 
The coupling was chosen to be $g_{\rm s}=0.1\,\Phi_-$.}\label{exponentHubble}
\end{figure}

So far we have restricted ourselves to tunneling in the flat Minkowski
background. 
In a FRW universe with scale factor $a$, the eigenvalue equation
(\ref{102}) changes 
to
\begin{eqnarray}\label{109}
\left(-\frac{d^2}{dt_E^2}-\frac{3\dot{a}}{a}\frac{d}{dt_E}+\frac{4\pi^2
    n^2}{L^2a^2}+2 g_{\rm s}\phi(t_E)\right)u(-\lambda,t_E)=\lambda\,
u(-\lambda,t_E)\,. 
\end{eqnarray}
With the ansatz
$u(-\lambda,t_E)=\varphi_1(-\lambda,t_E)\varphi_2(t_E)$ and an appropriate  
choice of $\phi_2(t_E)$, we eliminate the first derivative in (\ref{109}),
\begin{eqnarray}\label{110}
& & \left(-\frac{d^2}{dt_E^2}+\frac{4\pi^2
    n^2}{L^2a^2}+\frac{9\dot{a}^2}{4a^2}+2 g_{\rm
    s}\varphi(t_E)+\frac{3}{2}\frac{d}{dt_E}
\left(\frac{\dot{a}}{a}\right)\right)\varphi_1(-\lambda,t_E)\nonumber\\
& & \;\;\;=\lambda\, \varphi_1(-\lambda,t_E)\,. 
\end{eqnarray}
The scale factor is in general a complex function of the Euclidean
time $t_E$ that leads
to complex eigenvalues of the differential operator.
Assuming that no eigenvalues lie on the negative real axis, we can still
apply the regularization method.
The functional determinant related to (\ref{110}) is obtained by
performing the substitutions
\begin{eqnarray}
L^2\rightarrow L^2a^2
\end{eqnarray}
and
\begin{eqnarray}
2 g_{\rm s}\phi(t_E)\rightarrow 2 g_{\rm
  s}\phi(t_E)+\frac{9\dot{a}^2}{4a^2}+
\frac{3}{2}\frac{d}{dt_E}\left(\frac{\dot{a}}{a}\right)\,. 
\end{eqnarray} 
We restrict ourselves to flat de Sitter space with $\dot{a}/a=-i
H=$ const and with a small Hubble parameter, 
that is, $T_0,L\ll 1/H$ and $H\ll \omega$.
Using (\ref{infinitesum}), we find a correction term of order
$g_{\rm s} H^2$, 
\begin{eqnarray}\label{112}
\Gamma&=&\Gamma_0 \exp\Bigg[\frac{g_{\rm s} L \phi_-
  T_0}{12}-\frac{g_{\rm s} L^2\phi_-}{8\pi}\nonumber\\ 
& &-g_{\rm s} H^2\phi_-\left(\frac{LT_0^3}{72}-\frac{3L^4}{128\pi}-
\frac{9 L^3  T_0}{32 \pi^2}\left(1-\ln\left(
    \frac{4\pi}{L}\right)\right)\right)\Bigg]\,, 
\end{eqnarray}
where we have omitted terms that can be neglected for large $T_0$.
\begin{figure}[ht]
\centering
\includegraphics{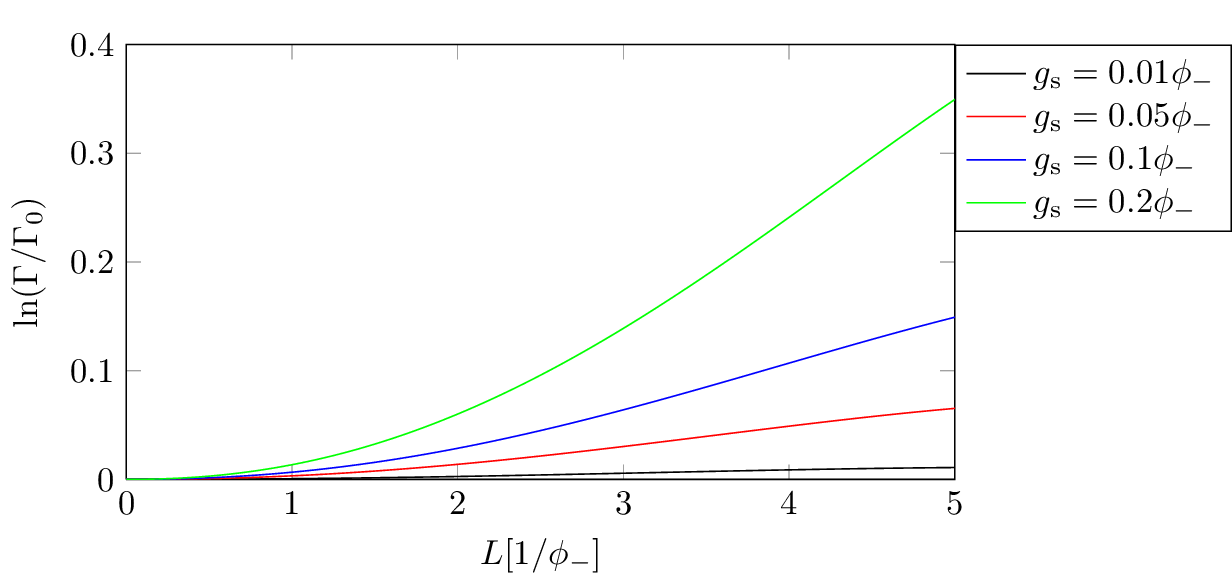}
\caption{We set $L=T_0$ for simplicity and show the change of the
  modified tunneling rate 
  with increasing $L$ for different couplings $g_s$.
(For this choice, the exponents in \eqref{108} and \eqref{112} still
remain positive.)}\label{exponent} 
\end{figure}
In order to discuss the result quantitatively, we depict in Fig.~\ref{exponent}
the dependence of $\ln(\Gamma/\Gamma_0)$ on the length $L$ which is
roughly the size of the nucleating vacuum bubble.
The Hubble parameter is set to zero, and we have evaluated
$\ln(\Gamma/\Gamma_0)$ using the exact expressions (\ref{106}) and (\ref{111}).
Obviously, the correction term of the nucleation rate increases with
$L$ and the coupling $g_{\rm s}$. 
In Fig.~\ref{exponentHubble} we depict the correction term of the
exponent in (\ref{112}) for small 
values of the Hubble parameter, that is, $L,T_0\ll1/H$.
The finite Hubble horizon leads to a reduction of the exponent for
small $L$ and $T_0$. 
We note that for unrealistically strong coupling one gets a
suppression of the tunneling rate (not shown in the figures); 
this is reminiscent of the quantum Zeno effect
\cite{deco,schlosshauer}. However,
in that limit is may no longer be realistic to neglect the back reaction on
the background. 

\section{Cosmic Landscape}

The cosmic landscape motivated by string theory was
discussed in various publications \cite{carr,Douglas}. 
Usually one considers Coleman--De~Luccia tunneling
\cite{coleman,coleman2,Lee,Linde1,Linde2} between a huge amount 
of vacua and discusses various solutions of ad hoc rate equations.
Under certain circumstances a continuum limit of these rate equations
can be derived \cite{Podolsky,Piao}. 

Furthermore, finite temperature effects have been considered in
\cite{Tye1} based on 
Hawking--Moss tunneling \cite{Hawking}.
Rapid tunneling was proposed under the assumption that resonance tunneling is
dominant in the landscape \cite{Sarangi,Tye4,Tye2}, 
see also \cite{copeland} for a critics based on standard quantum field
theory. 
In the following, we want to extend our model discussed in the
preceding sections 
to multi-level systems in order to see under which circumstances an
{\it ad hoc} rate 
equation can be formulated.

One can, for example, generalize the Hamiltonian (\ref{intsigma}) in
the following way: 
\begin{eqnarray}
H_\phi&=&H_\mathrm{diag}+H_\Delta= {\rm diag}
(\omega_1,...,\omega_n)+\sum_{i\neq j} \Delta_{ij}|i\rangle \langle
j|\ ,\\
H_\mathrm{int}&=& g_{\rm s} a^3(t) {\rm diag}(S_{11},...,S_{nn})\int
d^3p \,\sigma({\mathbf p})\sigma(-{\mathbf p})\ .\label{bathmediated} 
\end{eqnarray}
The interpretation is as follows: The numbers $\omega_i$ denote the
different local vacua 
of a cosmic landscape, and the  $\Delta_{ij}$ are tunneling matrix
elements which can  
be computed in WKB-approximation.
The entries $S_{ii}$ in the interaction Hamiltonian 
distinguish the different vacua, which is 
an obvious generalization of measuring the left and the right well in
the double-well system discussed above.

Since the tunneling matrix elements are usually exponentially small,
the short-time dynamics 
(short with respect to the tunneling times $\Delta^{-1}_{ij}$) is
determined by the decoherence rates. 
The off-diagonal elements of the density matrix now read, cf. \eqref{nondiag},
\begin{eqnarray}\label{offdiag}
\rho_{ij}(t)=\rho_{ij}(0)\exp\left(-\frac{g_{\rm
      s}^2(S_{ii}-S_{jj})^2}{4}\mathrm{Tr}
\left(\frac{(\Re\Omega^{'})^2+(\Im\Omega^{'})^2}{(\Re\Omega)^2}\right)
-i\varphi_{ij}\right)\   .  
\end{eqnarray}
One can conclude from this expression that the suppression of interference terms
depends crucially on the distance between different minima in the
landscape.

If one neglects possible degeneracies and assumes that the typical
decoherence rate 
is much larger than the tunneling rate, the system dynamics is
determined by the equations \cite{schlosshauer}
\begin{eqnarray}
\dot{\rho}_{ii}(t)=-\int_{0}^t ds[H_\Delta(t),[H_\Delta(s),\rho(s)]]_{ii}\,.
\end{eqnarray}
Applying the Markov approximation, one obtains
\begin{eqnarray}\label{markov}
\dot{\rho}_{ii}(t)=\lambda\sum_{k\neq i}|\Delta_{ik}|^2(\rho_{kk}-\rho_{ii})\,,
\end{eqnarray}
where $\lambda$ is chosen such that the coarse-graining in time is
not too small and the approximation is valid. 

There exists another generalization of the model where the tunneling
between different vacua may, in fact, be 
{\em mediated} by the environment. This is described by setting
$\Delta_{ij}=0$ and $S_{ij}\neq 0,i\neq j$ in the above Hamiltonian. 
Indirect coupling between metastable states is a well-known phenomena
in glasses, see, for example, \cite{Schechter}.

In the following we shall derive the master equations for the
environment-mediated tunneling.
The interaction Hamiltonian in the interaction picture has the form
\begin{eqnarray}
H^{I}_{\rm int}(t)= g_{\rm s} a^3(t)
\sum_{ij}e^{i(\omega_i-\omega_j)}S_{ij}|i\rangle \langle j|\int 
d^3p\, \sigma(\mathbf{p},t)\sigma(-\mathbf{p},t)\,.
\end{eqnarray}
Restricting ourselves to flat slices through de Sitter space, the
operators $\sigma(\mathbf{p},t)$ are given by 
\begin{eqnarray}
\sigma(\mathbf{p},t)=f_p(t) a_\mathbf{p}e^{i\mathbf{px}}+{\rm h.c.}
\end{eqnarray}
with \cite{Parker}
\begin{eqnarray}
f_p(t)=\sqrt{\frac{V}{(2\pi)^3}}\frac{1}{\sqrt{2p^3}}\left(\frac{p}{a(t)}+iH\right)e^{i\frac{p}{a(t)H}}\,.
\end{eqnarray}
Applying the Redfield approximation \cite{schlosshauer}, we find for
the system density matrix the expression
\begin{eqnarray}
\dot{\rho}_S^I(t)=-\mathrm{Tr}_B\int^t_0 ds [H^{I}_{\rm
  int}(t),[H^{I}_{\rm int}(s),\rho^I_S(t)\rho_B]]\,. 
\end{eqnarray}
In the limit of vanishing temperature, the bath density matrix is just
$\rho_B=|0\rangle\langle 0|$. 
The coefficients of the density matrix satisfy the system of
differential equations \cite{louisell}, 
\begin{eqnarray}\label{master}
\dot{\rho}_{ji}^I=\delta_{ji}\sum_{k\neq
  i}\rho_{kk}^I(w^+_{kiik}+w^-_{kiik})-\rho_{ji}^I 
\left[\sum_{l}(w^+_{jllj}+w^-_{illi})-w^+_{iijj}-w^-_{iijj}\right]
\end{eqnarray}
with the correlation functions
\begin{eqnarray}
w^+_{klmn}(t)&=&g_{\rm s}^2 \int_0^t ds
\,e^{i(\omega_{k}-\omega_{l})(s-t)}S_{kl}S_{mn}a^3(t)a^3(s)\nonumber\\ 
& &\times\int d^3p\int d^3q
\langle\sigma(\mathbf{p},s)\sigma(-\mathbf{p},s)
\sigma(\mathbf{q},t)\sigma(-\mathbf{q},t)\rangle  
\end{eqnarray}
and 
\begin{eqnarray}
w^-_{mnkl}(t)&=&g_{\rm s}^2 \int_0^t ds\,
e^{i(\omega_{k}-\omega_{l})(s-t)}S_{kl}S_{mn}a^3(t)a^3(s)\nonumber\\ 
& &\times\int d^3p\int d^3q
\langle\sigma(\mathbf{p},t)\sigma(-\mathbf{p},t)
\sigma(\mathbf{q},s)\sigma(-\mathbf{q},s)\rangle\,. 
\end{eqnarray}
In deriving (\ref{master}), several approximations have been performed:
the Born approximation, which states that the total density matrix can
be written 
approximately as a tensor product of the bath density matrix and the
system density matrix, and
the rotating wave approximation, which is valid when the intrinsic time
scale of the 
system is much larger than the relaxation time of the open system.
Since the correlation functions are not homogeneous in time due to the scale
factor, the master equation is not Markovian.

The rates in (\ref{master}) need not be exponentially small and may
therefore dominate the dynamics of the string landscape. 
The transition probabilities between the vacua are symmetric, since we
assume that the environment 
is described by a Gaussian wave function rather than an ensemble of states.
On the other hand, the tunneling rates in the Pauli equations 
(\ref{master}) are not symmetric and jumping 
to lower energy levels is more probable than jumping to higher energy levels,
depending on the bath temperature.
It would be interesting to see how the situation changes if the
Gaussian is replaced by a (micro)canonical ensemble.

Evaluating the correlation functions we find
\begin{eqnarray}\label{corr1}
& & w^+_{klmn}(t)=g_{\rm s}^2 \int_0^t ds\,
e^{i(\omega_{k}-\omega_{l})(s-t)}S_{kl}S_{mn}a^3(t)a^3(s) \nonumber \\ 
& &\; \times \frac{(2\pi)^3}{V}\int
d^3k\left(2(f_k(t)f^*_k(s))^2+\frac{V}{(2\pi)^3}\int
  d^3p|f_k(t)|^2|f_p(s)|^2\right) 
\end{eqnarray}
and
\begin{eqnarray}\label{corr2}
& & w^-_{mnkl}(t)=g_{\rm s}^2 \int_0^t ds\,
e^{i(\omega_{k}-\omega_{l})(s-t)}S_{kl}S_{mn}a^3(t)a^3(s)\nonumber \\ 
& &\; \times \frac{(2\pi)^3}{V}\int
d^3k\left(2(f_k(s)f^*_k(t))^2+\frac{V}{(2\pi)^3}\int
  d^3p|f_k(s)|^2|f_p(t)|^2\right)\,. 
\end{eqnarray}
The dominating contributions in the correlators (\ref{corr1}) and
(\ref{corr2}) are given by 
infrared contributions $k<Ha$, since the phases in the integrands are
oscillating 
rapidly if $k>Ha$.
Neglecting the second term in the parenthesis of (\ref{corr1}),
respectively (\ref{corr2}), and 
applying the approximation $e^{ik/Ha}\approx 1$ leads to
\begin{eqnarray}
w^+_{klmn}(t)&=&g_{\rm s}^2 \int_0^t ds\,
e^{i(\omega_{k}-\omega_{l})(s-t)}S_{kl}S_{mn}a^3(t)a^3(s)\\ 
& &\times\frac{2V}{(2\pi)^3}\int
d^3k\frac{1}{(2k^3)^2}\left(\frac{k}{a(t)}+iH\right)^2\left(\frac{k}{a(s)}-iH\right)^2   
\end{eqnarray}
and an analogous expression for $w^-_{mnkl}(t)$.
The evolution equation for the off-diagonal elements read in the
Schr\"odinger picture 
\begin{eqnarray}\label{offdiag2}
\dot{\rho}^\mathrm{Sch}_{ij}=-(\Gamma_{ij}-i(\omega_{i}-\omega_{j}))\rho_{ij}^\mathrm{Sch} 
\end{eqnarray}
with
\begin{eqnarray}
\Gamma_{ij}=\Re\left(\sum_{l}(w^+_{jllj}+w^-_{illi})-w^+_{iijj}-w^-_{iijj}\right)\,.
\end{eqnarray}
The imaginary part of the correlation function has been absorbed into
the frequencies $\omega_i$. 
For $S_{ij}=S_{ii}\delta_{ij}$ we obtain for $a(t)\gg a(0)$
\begin{eqnarray}\label{Gamma}
\Gamma_{ij}\approx g_{\rm s}^2
(S_{ii}-S_{jj})^2\frac{VH^3a^6(t)}{9(2\pi)^2}\left(\frac{1}{k_\mathrm{min}^3}-\frac{1}{k_\mathrm{max}^3}\right)\ , 
\end{eqnarray}
where $k_\mathrm{min}$ is an infrared cutoff and $k_\mathrm{max}\sim Ha$.
Solving (\ref{offdiag}) using the rates (\ref{Gamma}) gives
for large times 
\begin{eqnarray}\label{dens2}
|\rho_{ij}(t)|&=&\rho_{ij}(0)\exp\left(-g_{\rm s}^2\int^t_{t_0} dt'
  \Gamma_{ij}(t')\right)\nonumber\\ 
& \approx&\exp\left(-\frac{g_{\rm s}^2(S_{ii}-S_{jj})^2}{4}\frac{V a^6
    H^2}{56\pi^2}\left(\frac{1}{k_\mathrm{min}^3}-\frac{1}{k_\mathrm{max}^3}\right)\right)\,.  
\end{eqnarray}
This result can be compared with the off-diagonal element (\ref{nondiag}).
Using the dominant contribution of the exponent given by
(\ref{superhubble}), we find  
\begin{eqnarray}\label{dens1}
|\rho_{\pm}(t)|=\rho_{\pm}(0)\exp\left(-g_{\rm
    s}^2\frac{(\phi_+-\phi_-)^2}{4}\frac{V a^6
    H^2}{56\pi^2}\left(\frac{1}{p_\mathrm{min}^3}-
\frac{1}{p_\mathrm{max}^3}\right)\right)\, , 
\end{eqnarray}
which coincides with \eqref{dens2} if one identifies $S_{ii}$
($S_{jj}$) with $\phi_+$ ($\phi_-$).

The transition between the different localized vacuum states
is given by the rate equation
\begin{eqnarray}\label{rate}
\dot{\rho}_{ii}^\mathrm{Sch}=-2\sum_{k\neq i}(\Re w_{ikki}^+
\rho_{ii}^\mathrm{Sch}-\Re w_{kiik}^+ \rho_{kk}^\mathrm{Sch})\,.
\end{eqnarray}
For $a(t)\gg a(0)$ we find
\begin{eqnarray}
\Re w_{kiik}^+=\Re w_{ikki}^+=\frac{ g_{\rm s}^2 |S_{ik}|^2 V
  a^6(t)}{(2\pi)^2}\frac{H^5}{9H^2+(\omega_i-\omega_k)^2}\left(\frac{1}{k_\mathrm{min}^3}-\frac{1}{k_\mathrm{max}^3}\right)\,. 
\end{eqnarray}
Depending on the physical situation, that is, depending on whether a
bath-induced coupling between different 
vacua or the tunneling dominates, (\ref{rate}) or
(\ref{markov}) describes the evolution of the cosmic landscape. 
 
This evolution is described by $n$ coupled ordinary differential
equations and can obviously not be solved for $n\sim 10^{500}$ vacua; 
therefore, further assumptions and approximations are necessary
in order to obtain some physical insight. 

The approximation of Markov equations by Fokker--Planck equations is,
for example, described in \cite{vanKampen} and can always 
be applied if there is some small expansion parameter, for example, the ratio
of the jumps between different vacua and the size of the tunneling landscape.
A large number of vacua motivates the transition from discrete
values $\rho_{ii}(t)$ to a function 
$\rho_{x}(t)$, where $x$ is a continuous coordinate in a smooth cosmic
landscape. 
If the landscape is one-dimensional, one might consider the following scenario:
There are probabilities to go to the left and to the right, which are
described by functions $\alpha(x')$ and $\beta(x')$ if the observer
is located at a position $x'$. 
These functions are the continuum limit next-neighbor transition
rates in (\ref{markov}), 
\begin{eqnarray}
\lambda|\Delta_{i,i+1}|^2=\beta_i\rightarrow\beta(x),\nonumber\\
\lambda|\Delta_{i,i-1}|^2=\alpha_i\rightarrow\alpha(x)\ ,
\end{eqnarray}
respectively (\ref{rate}),
\begin{eqnarray}
2\Re w^+_{i,i+1,i+1,i}=\beta_i\rightarrow\beta(x),\nonumber\\
2\Re w^+_{i,i-1,i-1,i}=\alpha_i\rightarrow\alpha(x)\,.
\end{eqnarray}
The probability that there is a local minimum in the potential between $x$ and $x+dx$ is $\Omega \gamma(x) dx$, where $\Omega$ denotes the size of the cosmic landscape.
Therefore, we take into account the tunneling rates and the distance
to ``nearest neighbors'' of local vacua. 

The transition from the sum in (\ref{rate}) to a continuous description
can be performed as follows:
\begin{eqnarray}
\sum_{k}w^+_{kiik}\rho_{kk}&=:&\sum_{k}w(i,k)\rho(k)\nonumber\\
&=&\int dx(\delta(x-k_1)+...+\delta(x-k_n))w(i,x)\rho(x)\nonumber\\
&\approx& \int dx \int dk f(x-k)w(i,x)\rho(x)\nonumber\\
&=&\int dx  \gamma(x)w(i,x)\rho(x)\,.
\end{eqnarray}
Following the treatment of \cite{vanKampen} and assuming detailed balance,
\begin{eqnarray}
\beta(x')\gamma(x)\rho(x')_\mathrm{stat}=\alpha(x')\gamma(x')\rho(x)_\mathrm{stat}\,,  
\end{eqnarray}
where $\rho(x)_\mathrm{stat}$ is some stationary distribution, the
Fokker--Planck equation of diffusion type holds: 
\begin{eqnarray}\label{fokkerplanck}
\frac{\partial \rho(x,t)}{\partial
  t}=\frac{2}{\Omega}\frac{\partial}{\partial x}\frac{1}{\gamma(x)}
\frac{\partial}{\partial x}\frac{\rho(x,t)}{\rho_\mathrm{stat}(x)}\,. 
\end{eqnarray}
This equation describes diffusion in an inhomogeneous medium, since
the rates do not prefer a special direction in the cosmic 
landscape, that is, the dynamics is a random walk in an inhomogeneous medium.
The drift term in (\ref{fokkerplanck}) is due to
inhomogeneities in the cosmic landscape and  
vanishes for $\gamma(x)={\rm const}$. 
In the following we will rescale the time such that the factor $2/\Omega$ is absorbed.
If the rates are time-dependent as in (\ref{rate}), the diffusion
equation acquires a time-dependent factor on the right-hand side. 

Let us illustrate the solution of (\ref{fokkerplanck}) in two simple examples.
If the pure tunneling given by (\ref{markov}) dominates and
$\gamma(x)=\lambda\Delta^2$, where $\Delta$ is a typical tunneling
rate, we obtain the usual solution for the diffusion equation,
\begin{eqnarray}
\rho(x,t)=\frac{1}{\sqrt{\pi\lambda \Delta^2
    t}}\exp\left(-\frac{x^2}{\lambda \Delta^2 t}\right) \ .
\end{eqnarray}
If the dynamics is environment-induced and given by (\ref{rate}), the
diffusion depends on the scale factor. 
For a scalar-field environment and assuming $a(t)=\exp(H t)$, the
result is approximately given by
\begin{eqnarray}
\rho(x,t)=\sqrt{\frac{H}{\pi D(a^6-1)}}\exp\left(-\frac{H
    x^2}{D(a^6-1)}\right)\ , 
\end{eqnarray}
with
\begin{eqnarray}
D=\frac{g_{\rm s}^2 L^3 S^2}{H^2}\, ,
\end{eqnarray}
where $S$ denotes a typical transition element $S_{ik}$ in the
interaction (\ref{bathmediated}). 
Therefore the diffusion process may become faster 
due to the growth of the scale factor.
This is, of course, only possible if the cosmic landscape and its
environment exchange enough energy 
to lift the scalar field from one local minima to another.

We have shown in our paper how a small positive value of the
cosmological constant could be justified through decoherence. A
realistic scenario enabling the calculation of the observed value can,
however, only be presented after a definite cosmic landscape for the
potential has been retrieved from a fundamental theory. 
\section{Conclusion and Outlook}

In our paper we have addressed the problem why the cosmological
constant (dark energy) has a small positive value instead of being
exactly zero. Yokoyama had suggested that this could be due to the
dark-energy field not being in its ground state (whose energy is
assumed to be zero) but in a localized state \cite{yoko}.
Using a quantum mechanical model with a double-well potential, we have
justified the localization of the dark-energy field in one of the
minima. The crucial mechanism for this is decoherence -- the
emergence of classical properties (here, a
localized state) by irreversible and ubiquitous interaction with
irrelevant degrees of freedom (``environment'') leading to quantum
entanglement between the dark-energy field and environment \cite{deco}. This
localization is similar to the emergence of a chiral state for sugar molecules.

More precisely, we have considered a Yukawa interaction between the dark-energy 
scalar field (quintessence) and 
environmental modes that can be interpreted as gravitational degrees
of freedom. 
This interaction induces a dynamical suppression of interference
terms connecting the two minima. Consequently, the ground state (which
is a superposition of the two localized states) will for the
dark-energy field evolve into an ensemble of localized states with an
effective energy greater than zero.
The decoherence factor is dominated
by low-frequency super-Hubble modes.
These modes become strongly entangled with the dark-energy field because
their effective  
coupling involves an additional factor of $Ha/p$ compared to 
sub-Hubble modes. 

Motivated by recent ideas in string theory, we have generalized our
model and have considered an arbitrary number of perturbative vacuum
states. Again, strong decoherence leads to the localization in a
particular potential well. But one can also obtain the interesting
limit of an environment-induced transition between different minima.
Within the Markov approximation, we have derived rate equations describing
the latter feature.
These transitions can dominate the time evolution if the coupling is
strong enough to transfer sufficient energy from the environment to
the scalar field in order to lift the latter out of the local
minima. 
In the continuum limit we have found the Fokker--Planck equation for the
distribution of  
the dark energy within the landscape of possible vacuum states. 

Another issue addressed in our paper is the change of the
tunneling rate induced by the coupling to the environment.
Is is known from various models involving bilinear couplings between
system and environment, for example from the Caldeira--Leggett model
\cite{caldlegg}, that 
decoherence can lead to the stabilization of metastable states.
This is analogous to the quantum Zeno effect, that is, the prevention
of a decay process due to the continuous monitoring by the environment.

We have, however, considered in our model a tri- instead of a bilinear coupling 
and have used a field-theoretical renormalization procedure instead of
an ad hoc chosen cutoff frequency.
For vacuum bubbles smaller or roughly equal to the Euclidean
nucleation time, that is, $L< 2\pi T_0/3$, 
we have found an enhancement of the nucleation rate instead of the
usual suppression, whereas for $L> 2\pi
T_0/3$ we have found a suppression. The enhancement is a consequence of
field theory and similar to the Casimir effect.

Quantum interaction with the environment gives a natural reason
for the localization of the dark-energy field in a potential well and,
consequently, for a small positive value of the cosmological
constant. As long as the precise form of the potential is not known,
its exact value can, however, not be computed.

The present work can be extended in various directions.
For example, it would be interesting to study the
localization process when 
the dark-energy field is spatially inhomogeneous, that is, when it
adopts different values in distinct spatial regions.
An additional difficulty would then be the inclusion of collisions
between different vacuum bubbles.

Another possible extension would be the application of these ideas to
the inflationary era in the early Universe, where the energy scale is
much higher. The principal mechanisms of localization and modification
of tunneling rates are the same, but the quantitative details are
different. From these details one should be able to learn whether or how
decoherence can lead to a metastable de~Sitter solution and how this
could affect the dynamics of the transition to non-inflationary eras following 
inflation including (re)heating.
We hope to return to some of these issues in future publications.

\section*{Acknowledgements}
C.K. is grateful to the Max Planck Institute for Gravitational Physics, 
Potsdam, for its kind hospitality while part of this work was done. F.Q. 
acknowledges support from the Bonn--Cologne Graduate School (BCGS). A.A.S. 
acknowledges the RESCEU hospitality at the last stage of this project. He 
was also partially supported by the Russian Foundation for Basic Research under 
grant 09-02-12417-ofi-m and by the German Science Foundation (DFG) under grant 
436 RUS 113/333/10-2. We thank Andrei Barvinsky, Alexander Kamenshchik, 
and Jun'ichi Yokoyama for 
helpful discussions and critical comments. 
\section*{Appendix}
In this section we shall give a short review of the
$\zeta$-function renormalization method as presented in
\cite{barkamkar}.
The authors there considered the renormalization of a functional determinant
defined by a second-order differential equation.
Usually, neither the eigenvalues nor the eigenfunctions of 
the differential operator are known exactly. 
Moreover, even if all the eigenvalues are known, the determinant is an
infinite product of eigenvalues, which is in general a divergent
quantity.

To solve these problems, one represents the functional determinant
via a generalized Riemann $\zeta$-function. 
The determinant of an arbitrary differential operator $\mathcal{D}$
can be written as 
\begin{eqnarray}\label{Det}
(\mathrm{Det}\mathcal{D})^{1/2}=\exp\left(\frac{1}{2}\ln\prod_\lambda
  \lambda\right)=\exp\left(\frac{1}{2}\sum_\lambda \ln
  \lambda\right)=\exp(W)\,, 
\end{eqnarray}
where the eigenvalues of the operator are denoted by $\lambda$.
We define the generalized $\zeta$-function through
\begin{eqnarray}\label{A2}
\zeta(s)=\sum_\lambda \frac{1}{\lambda^s}\,,
\end{eqnarray}
which is a convergent series for some $s>0$ and can be continued
analytically to $s=0$.
The exponent $W$ in equation (\ref{Det}) can be obtained through
\begin{eqnarray}\label{A1}
W=-\frac{1}{2}\frac{d}{ds}\zeta(s)\Bigg|_{s=0}\,.
\end{eqnarray}
Since the eigenvalues of $\mathcal{D}$ have the dimension of mass
squared (recall that $\hbar=1$), this leads to a wrong dimensionality for $W$.
Therefore we have to replace (\ref{A1}) by 
\begin{eqnarray}\label{A3}
W=-\frac{1}{2}\frac{d}{ds}\sum_\lambda
\left(\frac{\mu^2}{\lambda}\right)^s
\Bigg|_{s=0}=-\frac{1}{2}\zeta'(0)-\frac{1}{2}\zeta(0)\ln\mu^2,   
\end{eqnarray}
where we have introduced a renormalization parameter $\mu$ with mass
dimension one. 

The differential operator corresponding to a single field mode may be
labelled by $n$. 
In quantum mechanics we are confronted with a \textit{finite} number
of modes, whereas in field theory we have to deal with an \textit{infinite}
number.
For each fixed $n$, the eigenvalue equation reads
\begin{eqnarray}
D_n u_n(-\lambda,t)=\lambda u_n(-\lambda,t)\ ,
\end{eqnarray}
where $\lambda$ is determined by the boundary condition
\begin{eqnarray}\label{A50}
u_n(-\lambda,t_0)=0\,.
\end{eqnarray}
This boundary condition together with a normalization determines the
eigenfunctions uniquely. 
All the boundary conditions (\ref{A50}) can be collected in the equation 
\begin{eqnarray}\label{A4}
\mathrm{Det} \, u_n(-\lambda,t_0)=0\ ,
\end{eqnarray}
where the determinant is taken with respect to all modes $n$ and all
eigenvalues $\lambda$. 
With the help of the Cauchy formula, the generalized $\zeta$-function can be expressed as 
\begin{eqnarray}
\zeta(s)=\frac{1}{2\pi i}\int_C
\frac{dz}{z^s}\frac{d}{dz}\sum_n\ln u_n(z,t_0)\ , 
\end{eqnarray}
with the contour $C$ encircling all roots of equation (\ref{A4}).
Deforming the contour $C$ to a contour $\tilde{C}$ which
encircles the branch cut of the function $z^{-s}$, we find
\begin{eqnarray}\label{A5}
\zeta(s)=\frac{\sin(\pi s)}{\pi}\int_0^\infty \frac{d
  M^2}{M^{2s}}\frac{d}{d M^2}\sum_n\ln u_n(M^2,t_0)\,.
\end{eqnarray}
We first consider the regularization method for a quantum mechanical system.
The necessary information for the regularization of a 
system with a finite number of modes is contained in the function
\begin{eqnarray}
I(M^2)=\sum_n \ln u_n(M^2,t_0)\,.
\end{eqnarray}
Expanding this function for large $M$ leads to
\begin{eqnarray}
I(M^2\rightarrow\infty)=\sum_{k=1}^{N}(I_k+\bar{I}_k\ln
M^2)M^{2k}+(I^\mathrm{R})_\mathrm{ln}\ln M^2+I^{R}(\infty)\,, 
\end{eqnarray}
where $(I^\mathrm{R})_\mathrm{ln}$ is the coefficient of the
logarithmic asymptotic term of $I(M^2)$, 
and $I^{R}(\infty)$ is the asymptotic value of the regular part of $I(M^2)$.
According to \cite{barkamkar}, the $\zeta$-function can be expanded as
\begin{eqnarray}\label{A51}
\zeta(s)=(I^\mathrm{R})_\mathrm{ln}+s[I^\mathrm{R}]^{\infty}_0+\mathcal{O}(s^2)\,,
\end{eqnarray}
where $[I^\mathrm{R}]^{\infty}_0=I^\mathrm{R}(\infty)-I^\mathrm{R}(0)$.
As a demonstration we will apply this method to the harmonic oscillator.
The eigenvalue equation
\begin{eqnarray}
\left(-\frac{d^2}{dt^2}+\omega^2\right)u(-\lambda,t)=\lambda
u(-\lambda,t)\,,\quad u(-\lambda,t_0)=0 \ ,
\end{eqnarray}
has a solution of the form
\begin{eqnarray}\label{harmosc}
u(-\lambda,t)=Ae^{\sqrt{\omega^2-\lambda}t}+Be^{-\sqrt{\omega^2-\lambda}t}\,.
\end{eqnarray}
Performing the analytical continuation to the complex plane,
$\lambda\rightarrow z$, 
the function $u$ adopts on the negative real axis the form
\begin{eqnarray}
 u(M^2,t)=Ae^{\sqrt{\omega^2+M^2}t}\,,
\end{eqnarray}
where we have neglected the exponentially decreasing term.
Using for convenience the normalization $u'(0)=1$ leads to 
\begin{eqnarray}\label{Iharmosc}
I(M^2)=-\frac{1}{2}\ln(\omega^2+M^2)+\sqrt{\omega^2+M^2}t\,.
\end{eqnarray}
Since the term proportional to $\exp(-\sqrt{\omega^2 +M^2}t)$ has been
neglected, 
the analytically continued eigenfunctions do not respect the boundary
condition $u(-\lambda,t_0)=0$. 

 From (\ref{Iharmosc}) we find $I^\mathrm{R}(0)=-\ln \omega+\omega t$,
$(I^\mathrm{R})_\mathrm{ln}=-1/2$. 
Using equations (\ref{Det}), (\ref{A3}), and (\ref{A51}) we arrive at
\begin{eqnarray}
\left(\mathrm{Det}\left[-\frac{d^2}{dt^2}+\omega^2\right]\right)^{-1/2}
=\sqrt{\frac{\omega}{\mu}}e^{-\frac{\omega t}{2}} 
\end{eqnarray}
which gives for large $t$ the correct ground-state energy $\omega/2$ for a
harmonic oscillator, cf. Eq. (2.16) in the second reference of
\cite{coleman}. 

In general, the exact shape of the eigenfunctions $u_n(-\lambda,t)$ is unknown
and one approximates the $u_n$ with a uniform asymptotic WKB-expansion.
This asymptotic expansion has the property that 
\begin{eqnarray}\label{WKBexp}
\ln u_n(M^2,t_0)=\phi_{\rm WKB}(n^2,M^2/n^2)
\end{eqnarray}
is uniform for $M^2/n^2\rightarrow\infty$ and $M^2/n^2\rightarrow0$.
In addition, it is also possible to use (\ref{WKBexp}) for the
regularization of functional determinants in field theory, that is, if
the mode number $n$ is not bounded.
The expansion (\ref{WKBexp}) has at most a \textit{finite} power-law order growth in $n$ \cite{barkamkar,olver}.
This fact allows us to use the parameter $s$ to cure the divergences
arising from the infinite number of modes. 
Changing the integration variable from $M^2\rightarrow n^2 M^2$ leads to
\begin{eqnarray}\label{WKBzeta}
\zeta(s)=\frac{\sin(\pi s)}{\pi}\int_0^\infty \frac{d
  M^2}{M^{2s}}\frac{d^2}{dM^2}I(M^2,s)\,
\end{eqnarray}
with
\begin{eqnarray}
I(M^2,s)=\sum_n \frac{1}{n^{2s}}\ln u_n(M^2 n^2,t_0)\,.
\end{eqnarray}
For a finite parameter $s>0$ the expression (\ref{WKBzeta}) is finite.
Analytic continuation of the $\zeta$-function from its convergence domain to $s=0$ leads to \cite{barkamkar}
\begin{eqnarray}\label{A6}
\zeta(s)&=&\frac{1}{s}(I^{\mathrm{pole}})_{\mathrm{ln}}+(I^\mathrm{R})_{\mathrm{ln}}+[I^{\mathrm{pole}}]^\infty_0\nonumber\\ 
& &+s\left\{[I^\mathrm{R}]^\infty_0-\int_0^\infty d M^2\ln M^2\frac{d
    I^{\mathrm{pole}}(M^2)}{d M^2}\right\}+\mathcal{O}(s^2) \,.
\end{eqnarray}
The coefficients $(I^{\mathrm{pole}})_{\mathrm{ln}}$, $I^{\mathrm{pole}}(\infty)$ and $I^\mathrm{R}(\infty)$
are defined through the large $M$-expansion
\begin{eqnarray}
I(M^2\rightarrow\infty,s)&=&\frac{(I^{\mathrm{pole}})_{\mathrm{ln}}\ln
  M^2+I^\mathrm{pole}(\infty)}{s}\nonumber\\
& &+I^\mathrm{R}(\infty)+(I^\mathrm{R})_{\mathrm{ln}}\ln M^2+\mathcal{O}(M^2)\ ,
\end{eqnarray}
and the pole part $I^{\mathrm{pole}}(M^2)$ is defined through
\begin{eqnarray}
I(M^2,s)=\frac{I^{\mathrm{pole}}(M^2)}{s}+\mathcal{O}(s^0)\,.
\end{eqnarray}
$I^{\mathrm{pole}}(0)$ and $I^\mathrm{R}(0)$ are determined by
\begin{eqnarray}
I(M^2\rightarrow0,s)=\frac{I^{\mathrm{pole}}(0)}{s}+I^\mathrm{R}(0)+\mathcal{O}(s)\,.
\end{eqnarray}

It is also possible to apply the regularization method if the
differential equation exhibits singular coefficients.
According to Olver \cite{olver}, the WKB expansion of a second-order
differential equation 
\begin{eqnarray}\label{A7}
\frac{d^2 u(M^2,t)}{d t^2}=\omega(t)^2 u(M^2,t)=[f(t)+g(t)]u(M^2,t)
\end{eqnarray}
has the form 
\begin{eqnarray}
u(M^2,t)=C(M)(g(t))^{-1/4} \exp\left\{\int_0^t d t'
  (g(t'))^{1/2}\right\}\left[1+h(t)\right] 
\end{eqnarray}
with $h(t)=\mathcal{O}(M^{-1})$.
The function $h(t)$ can be expressed as Volterra integral
\begin{eqnarray}\label{A8}
h(t)&=&\frac{1}{2}\int^t_{t_0}\left(1-\exp\left\{2\int^{t'}_0
    g^{1/2}(t'')dt''-2\int^{t}_0g^{1/2}(t')dt'\right\}\right)\nonumber\times\\ 
& &\times\psi(t')[1+h(t')]dt'
\end{eqnarray}
with
\begin{eqnarray}
\psi(t)=\frac{f(t)}{g^{1/2}(t)}-\frac{1}{g^{1/4}(t)}\frac{d^2}{dt^2}\frac{1}{g^{1/4}(t)}\,.   
\end{eqnarray}
Therefore the WKB expansion only makes sense if the kernel of 
(\ref{A8}) is bounded. This leads to the condition  
\begin{eqnarray}
\Psi(t)=\int_{t_0}^{t}dt' |\psi(t')|<\infty\ .
\end{eqnarray}
Here, $t_0$ and $t$ are the boundaries of the interval under consideration.
The split of $\omega(t)^2$, see (\ref{A7}), is chosen such that singular coefficients like $1/t$ in the
differential equation do not destroy the WKB expansion. 
This is the reason for the $1/4$--trick used in \cite{barkamkar}.

As already mentioned above, the evaluation of $I(M^2,s)$ involves 
an important approximation. 
In order to fulfill the boundary condition (\ref{A50}), two 
linearly independent solutions of the corresponding differential
equation are required. 
After analytical continuation, the solutions are of the form
$u\sim\exp(Mt)$ and $u\sim\exp(-Mt)$. 
The second solution is exponentially decreasing and will therefore be discarded.
This implies that the analytically continued functions $u(M^2,t_0)$
\textit{do not} respect the boundary condition (\ref{A50}).

\end{document}